\documentclass[onecolumn,authoryear]{els-mrw}

\usepackage{amsmath,amssymb,amsfonts,amsthm,makeidx,graphicx}
\usepackage{txfonts}
\usepackage{ifthen}
\usepackage{wrapfig}
\usepackage{graphicx}
\usepackage{helvet}
\usepackage{derivative}
\usepackage{mathtools}
\usepackage[thinc]{esdiff}

\newcommand{\poltens}{\epsilon}
\newcommand{\introref}[1]{\noindent\textit{{#1}}}

\newcommand{\hdfunc}{\mu_\trt{HD}}
\newcommand{\nHz}{\tr{nHz}}
\newcommand{\hsn}{h_{s,n}}
\newcommand{\aisco}{a_\trt{ISCO}}
\newcommand{\fisco}{f_\trt{ISCO}}
\newcommand{\hscirc}{h_{s,\tr{circ}}}
\newcommand{\frstorb}{f_{r,\trt{orb}}}
\newcommand{\lgwcirc}{L_\tr{circ}}

\newcommand{\fgw}{f_\trt{GW}}
\newcommand{\mbh}{M_\textrm{BH}}
\newcommand{\fbulge}{f_\mathrm{bulge}}
\newcommand{\mbulge}{M_\mathrm{bulge}}
\newcommand{\mmbulge}{{$\mbh$--$\mbulge$}}
\newcommand{\sigmastar}{$\sigma_\star$}
\newcommand{\msigma}{{$M_\tr{BH}$--\sigmastar}}

\newcommand{\tobs}{T_\tr{obs}}
\newcommand{\taufgw}{\tau_{f,\trt{GW}}}
\newcommand{\tauegw}{\tau_{e,\trt{GW}}}
\newcommand{\taulife}{\tau_\tr{life}}
\newcommand{\taulifegw}{\tau_{\tr{life},\trt{GW}}}
\newcommand{\dtobs}{{\Delta T}_\tr{cad}}
\newcommand{\sigmatoa}{\sigma_{\Delta t}}
\newcommand{\sigmaradio}{\sigma_\tr{radio}}
\newcommand{\distcoord}{L_c}
\newcommand{\energydensgw}{\mathcal{E}_\tr{gw}}
\newcommand{\energygw}{E_\tr{gw}}
\newcommand{\lumgw}{L_\tr{gw}}

\definecolor{purple1}{rgb}{0.6, 0.0, 0.8}
\definecolor{green1}{rgb}{0.25, 0.5, 0.25}
\definecolor{red1}{rgb}{0.7, 0.15, 0.15}

\newcommand*{\needcite}[1]{
    \ifthenelse{\equal{#1}{}}{
        {\color{red1}[?]}
    }{
        {\color{red1}[#1]}
    }
}

\newcommand{\tr}[1]{\textrm{#1}}
\newcommand{\trt}[1]{\textrm{\tiny{#1}}}

\newcommand{\s}{\mathrm{s}}
\renewcommand{\sec}{\mathrm{s}}
\newcommand{\ns}{\mathrm{ns}}

\newcommand{\MHz}{\mathrm{MHz}}
\newcommand{\ms}{\mathrm{ms}}
\newcommand{\m}{\mathrm{m}}
\newcommand{\cm}{\mathrm{cm}}

\newcommand{\kg}{\mathrm{kg}}
\newcommand{\msol}{\tr{M}_{\odot}}

\newcommand{\kpc}{\mathrm{kpc}}
\newcommand{\pc}{\mathrm{pc}}
\newcommand{\yr}{\mathrm{yr}}        % yr in math-mode
\newcommand{\pyr}{{\textrm{yr}^{-1}}}

\newcommand{\model}{\Upsilon}
\newcommand{\gmn}{g_{\mu\nu}}
\newcommand{\nsamp}{N_\tr{samp}}
\newcommand{\covgw}{H}
\newcommand{\covnoise}{N}
\newcommand{\ostat}{S_\tr{opt}}

\newcommand{\dt}{{\tau}}
\newcommand{\Dt}{{\Delta{}t}}
\newcommand{\etamn}{\eta_{\mu\nu}}

\newcommand{\hmn}{h_{\mu\nu}}
\newcommand{\hij}{h_{ij}}
\newcommand{\hplus}{h_{+}}

\newcommand{\eplus}{\poltens^{+}}
\newcommand{\ecross}{\poltens^{\times}}

\newcommand{\comdist}{d_\tr{com}}
\newcommand{\fcoal}{f_\tr{coal}}
\newcommand{\fagn}{f_\trt{AGN}}
\newcommand{\mchirp}{\mathcal{M}}     % Chirp-mass

\newcommand{\E}[1]{\times\nobreak10^{#1}}

\newcommand{\logten}[1]{\log_{10}\!\lr{#1}}

\newcommand{\cospar}[2][]{\cos^{#1}\!\lr{#2}}
\newcommand{\msp}{\;\;\;}

\newcommand{\lr}[2][]{
    \ifthenelse{\equal{#1}{}}{
        {\left(#2\right)}
    }{
        {\left(#2\right)}^{#1}
    }
}

\newcommand{\lrs}[2][]{
    \ifthenelse{\equal{#1}{}}{
        {\left[#2\right]}
    }{
        {\left[#2\right]}^{#1}
    }
}

\newcommand{\scale}[3][]{
    \ifthenelse{\equal{#1}{}}{
        \lr{ \frac{#2}{#3} }
    }{
        {\lr[#1]{ \frac{#2}{#3} }}
    }
}
\newcommand{\scales}[3][]{
    \ifthenelse{\equal{#1}{}}{
        \lrs{ \frac{#2}{#3} }
    }{
        {\lrs[#1]{ \frac{#2}{#3} }}
    }
}

\usepackage{hyperref}
\hypersetup{
    colorlinks=true,
    linkcolor=blue,
    filecolor=magenta,
    urlcolor=cyan,
    pdfpagemode=FullScreen,
}

\renewcommand{\odv}[3][]{
    \ifthenelse{\equal{#1}{}}{
        \frac{\partial{}{{#2}}}{\partial{}{{#3}}}
    }{
        \frac{\partial{}^{#1}{{#2}}}{\partial{}{{#3}}^{#1}}
    }
}

\begin{document}

\chapter{Pulsar Timing Arrays}\label{PTAs}

\author[1]{Luke Zoltan Kelley}

\address[1]{\orgname{University of California, Berkeley}, \orgdiv{Department of Astronomy}, \orgaddress{501 Campbell Hall, Berkeley, CA 94720-3411}}

\maketitle

\setlength{\intextsep}{12pt}
\setlength{\columnsep}{10pt}

\begin{abstract}[Abstract]
    Pulsar Timing Arrays (PTAs) have recently found strong evidence for low-frequency gravitational waves (GWs) in the nanohertz frequency regime.  As GWs pass, they produce deviations in measured lengths and light-travel times.  PTA experiments utilize the highly-consistent radio bursts from millisecond pulsars, distributed throughout the local galaxy, to identify miniscule timing deviations indicative of GWs.
    To distinguish GWs from noise, PTAs search for a particular correlation pattern between different pulsars called Hellings~\&~Downs correlations.
    The type of GW signal that has recently been identified is a stochastic GW background (GWB), which is observed to have more power at lower GW frequencies.  A GWB matching these observations has long been predicted from super-massive black-hole (SMBH) binaries.  SMBHs are known to exist in the centers of galaxies, which can then form binaries when two SMBHs are brought together following the merger of galaxies.  No example of an SMBH binary has confidently been identified to date, and tremendous uncertainties about their formation and evolution remain.  Alternative sources of the GWB have also been proposed, based on models for new fundamental physics, particularly in the early Universe.  Improved sensitivity of PTAs will eventual lead to the characterization of GWB anisotropy and constraints on GWs from individual SMBH binaries, either of which could definitively demonstrate the true origin of the GWB.  If the source is SMBH binaries, a variety of electromagnetic counterparts are possible, allowing for multimessenger astrophysics with low-frequency GWs.
\end{abstract}


\begin{BoxTypeA}{Key Concepts}
\begin{itemize}
    \item Gravitational waves (GWs) are traveling perturbations in space-time that carry energy and momentum from sources, such as binaries, to distant observers.  Binaries from super-massive black-holes (SMBHs) can produce very strong gravitational waves in the low-frequency (nanohertz) regime, both as `continuous waves' (CWs) from individual binaries, and a stochastic `gravitational wave background' (GWB) from an ensemble of large numbers of binaries.
    \item GWs can be detected by identifying correlations in the patterns of length of time changes, called strain, ($h = \Delta T / T = \Delta L / L$).  Pulsar Timing Arrays (PTAs) examine correlations in the arrival times of radio bursts from millisecond pulsars distributed throughout the local galaxy.  The characteristic correlation pattern for PTAs is the Hellings~\&~Downs curve, which describes how similar the signals should be in any two different pulsars across the sky, based only on the angle between them.
    \item Extracting GW signals from PTA data requires careful modeling of many complex noise processes including motion of the pulsar and the solar system, and relativistic effects at each; dispersion and scattering from the interstellar medium; and noise/spin-evolution intrinsic to each pulsar.  Careful statistical analyses which incorporate our noise models and the expected Hellings~\&~Downs correlations are able to decipher GWB amplitudes as low as $\sim 10^{-15}$.
    \item Recently, PTAs including NANOGrav and many others across the globe, have identified strong evidence for a GWB signal consistent with the predictions from populations of SMBH binaries (SMBHBs).  Additional data will soon confirm if the signal is truly from GWs; and, if so, determine whether or not the GWB is indeed produced by SMBHBs or instead from new, `beyond the Standard Model' physics.  The `GWB spectrum' (strain~vs.~frequency curve) can encode significant information about SMBHB populations and evolution.
    \item SMBH binaries are formed following the merger of galaxies, each of which are known to contain a central SMBH with properties closely correlated with their host galaxy.  Environmental interactions between binaries and their host galaxies are always required for binaries to reach the small separations at which nanohertz GWs are produced.  SMBHB evolution is highly uncertain, but dynamical friction and stellar scattering are always required, while circumbinary accretion disks and a number of other processes could also be important at times.
    \item In addition to GWs, SMBHBs which are actively accreting can produce bright active galactic nucleus (AGN) emission across the electromagnetic (EM) spectrum.  GW emitting binaries cannot typically be resolved in images, but their binary motion can be encoded in a variety of time- or frequency- dependent EM signatures.  While many candidates have been identified, no SMBHBs have been confirmed.  Multi-messenger observations combining GWs and EM emission offer tremendous promise in understanding SMBHB formation and evolution.
\end{itemize}
\end{BoxTypeA}

\begin{glossary}[Glossary]
    \term{Binary Hardening}.  `Hardening' refers to the process of binaries gradually shrinking their separation due to the loss of orbital energy.  For massive black-hole binaries, this can result from gravitational wave emission, or `environmental interactions' between the binary and host galaxy. See Secs.~\ref{sec:gw-power-evolution}~\&~\ref{sec:binary_evolution}. \\
    \term{Massive Black Hole / Super-Massive Black Hole (SMBH) / Super-Massive Black-Hole Binary (SMBHB)}.  BHs are the massive remnants of objects whose extreme gravitational force leads to indefinite collapse, and the formation of an `event horizon' from which nothing can escape.  SMBHs have masses of $10^5 - 10^{10}~\msol$ and are observed to reside in the centers of galaxies.  SMBHBs can form following the merger of two galaxies, when the SMBHs become gravitationally bound to each-other in the post-merger galaxy.  See Secs.~\ref{sec:intro}~\&~\ref{sec:mbhbs}. \\
    \term{Burst With Memory}. A GW signature produced at the final coalescence of binaries due to a near-discontinuity, or `DC offset', in the space-time metric from before to after coalescence. See Sec.~\ref{sec:coalescence}. \\
    \term{Chirp Mass}.  A quantity calculated from the two masses of a binary (Eq.~\ref{eq:chirp_mass}) which directly determines the amplitude of binary GW signals, and the rate of binary inspiral due to GW emission. \\
    \term{Continuous Waves (CWs)}.  Nearly monochromatic GWs produced by individual binaries, where the GW strain pattern is roughly sinusoidal in time.  Contrasted with a `GWB'.  See Sec.~\ref{sec:binary_gws}. \\
    \term{Detection Statistic}. A quantity calculated from data which can be used as a quantitative measure of how confidently a signal can be detected in the data. See Sec.~\ref{sec:os}. \\
    \term{Environmental Interactions}.  Dynamical processes between an SMBHB and its local galactic environment which serve to change (typically extract) orbital energy, leading to changes in the binary separation (typically shrinking it).  See Sec.~\ref{sec:binary_evolution}. \\
    \term{Final-Parsec Problem}.  The possible stalling of SMBHB inspiral at parsec-scale separations, due to the depletion of stars that participate in stellar scattering, called `loss-cone' stars.  See Sec.~\ref{sec:binary_evolution_scattering}. \\
    \term{Gravitational Wave (GW)} a self-sustained perturbation of spacetime that carries energy and momentum, and produces measurable deviations in time and distance.  See Sec.~\ref{sec:gws_gr-waves}. \\
    \term{Gravitational Wave Background (GWB) / GWB Spectrum}.  The stochastic summation of GWs from large numbers of individual binaries each emitting at different frequencies.  The GWB is typically described by a characteristic strain spectrum over frequencies $h_c(f)$, or a power spectral-density $S_h(f)$.  See Sec.~\ref{sec:gws_gwb}. \\
    \term{GW Strain / Characteristic Strain.} Strain is the dimensionless fractional deviation in proper/measured distances and light-travel times, $h \approx \Delta L / L \approx \Delta T / T$, which is produced by passing GWs.  Characteristic strain is a measure of strain that can be used in signal-to-noise ratio calculations (Eqs.~\ref{eq:characterist_strain_mono}~\&~\ref{eq:characterist_strain_chirp}).  See Sec.~\ref{sec:gws_strain}. \\
    \term{Hellings \& Downs (HD) correlations.}  The spatial correlation pattern between GW detectors at different angles, when the light-travel distance is comparable or longer than the GW wavelength, such as in PTAs (Eqs.~\ref{eq:hd}). \\
    \term{Pulsar Timing Array (PTA)}. A collection of pulsars whose timing residuals are cross-correlated against each-other to search for correlated signatures such as GWs.  See Sec.~\ref{sec:ptas}. \\
    \term{Times of Arrival (TOAs) / Timing Residuals.}  TOAs are the measured times at which radio pulses from pulsars are measured by an observer.  Timing residuals are the difference between observed TOAs and the predictions/expectations from a timing model (Eqs.~\ref{eq:residual-timing-model}~\&~\ref{eq:toa-components-gw}).  See Secs.~\ref{sec:pulsar-timing}~\&~\ref{sec:os}. \\
    \term{Timing Model.} A model that predicts the pulse TOAs from a pulsar, typically accounting for a variety of `noise' processes which can modify the TOAs.  See Sec.~\ref{sec:pulsar-timing}~\&~\ref{sec:noise-sources}.  \\
    \end{glossary}

    \begin{glossary}[Units \& Constants]
    \begin{tabular}{@{}lp{34pc}@{}}
    $\msol$ & Solar mass, unit of mass, $1.9884\E{30} \, \kg$ \\
    $\pc$ & Parsec, unit of distance, $3.0857\E{16} \, \m$ \\
    $G$ & Newton's gravitational constant, $6.6743\E{{-11}} \, \m^3 \, \, \kg^{-1} \s^{-2}$ \\
    $c$ & Speed of light in a vacuum, $2.998\E{8} \, \m \, \s^{-1}$ \\
    $H_0$ & Hubble constant at redshift zero, $\approx 70 \, \tr{km}~\tr{s}^{-1}~\tr{Mpc}^{-1} \approx 2\E{-18} \, \s^{-1}$.
    \end{tabular}
    \end{glossary}

% ==============================================================================
% ====    Content   ====
% ==============================================================================

% ---- Introduction
% ------------------------------------------------------------------------------

\section{Introduction and Overview}\label{sec:intro}

Einstein's theory of general relativity provides a purely-geometric description of spacetime and gravity, and a means for reconciling measurements made from different reference frames.  After this theory developed, it was quickly realized that it implied two, nearly-incredible possibilities.  First, that objects could exist which are so compact that no force is able to prevent their indefinite collapse, producing regions within which nothing can escape.  Second, that the geometric structure allows for traveling perturbations in the shape of spacetime that carry energy to large distances.  Evidence for the first objects, called \term{black holes}, grew consistently after the early 20th century, culminating in the first direct images of black holes in 2019 by the Event Horizon Telescope.  Similarly, the second phenomenon, called \term{gravitational waves (GWs)}, were indirectly measured in the 1980s, and directly detected in 2015 by the LIGO-Virgo experiments.

\term{Pulsar Timing Arrays (PTAs)} are experiments to detect GWs in a new regime: at very-low frequencies.  While LIGO-Virgo is sensitive to kilohertz GW frequencies, with thousands of oscillations per second, PTAs are sensitive to nanohertz (nHz) frequencies: with oscillations only every few years.  The underlying principle behind detecting GWs is mostly the same between these two very-different frequency regimes.  As GWs pass by the observer and their measurement devices, distances in spacetime are modulated in predictable patterns.  By very precisely measuring deviations in light-travel times, and comparing these deviations between different places, those patterns of length variations can be measured and compared to GW predictions.

LIGO-Virgo, now with KAGRA, use laser beams running along two, perpendicular kilometers-long laser beams which are interfered together to produce a null signal in the absence of GWs.  As GWs pass, the two laser beams traverse differing distances which leads to the interference becoming imperfect, and a signal is then measurable by photo-detectors.  In these `laser interferometers', noise sources (e.g.~seismic vibrations) also produce imperfect interference.  By comparing, or `correlating', the signals between detectors at different locations across the Earth, noise can be filtered out, and subtle GW signals can be detected.  The amplitude of GWs is typically characterized in terms of the GW \term{strain}: the fractional change in length over some baseline distance, or fractional changes in light-travel time, $h = \Delta L / L = \Delta T / T$ (Sec.~\ref{sec:gws}).  Terrestrial laser interferometers are able to achieve strain sensitivities on the order of $h \sim 10^{-24}$.

\begin{wrapfigure}{r}{0.45\textwidth}
    \vspace{-6pt}
    \includegraphics[width=0.45\textwidth]{{{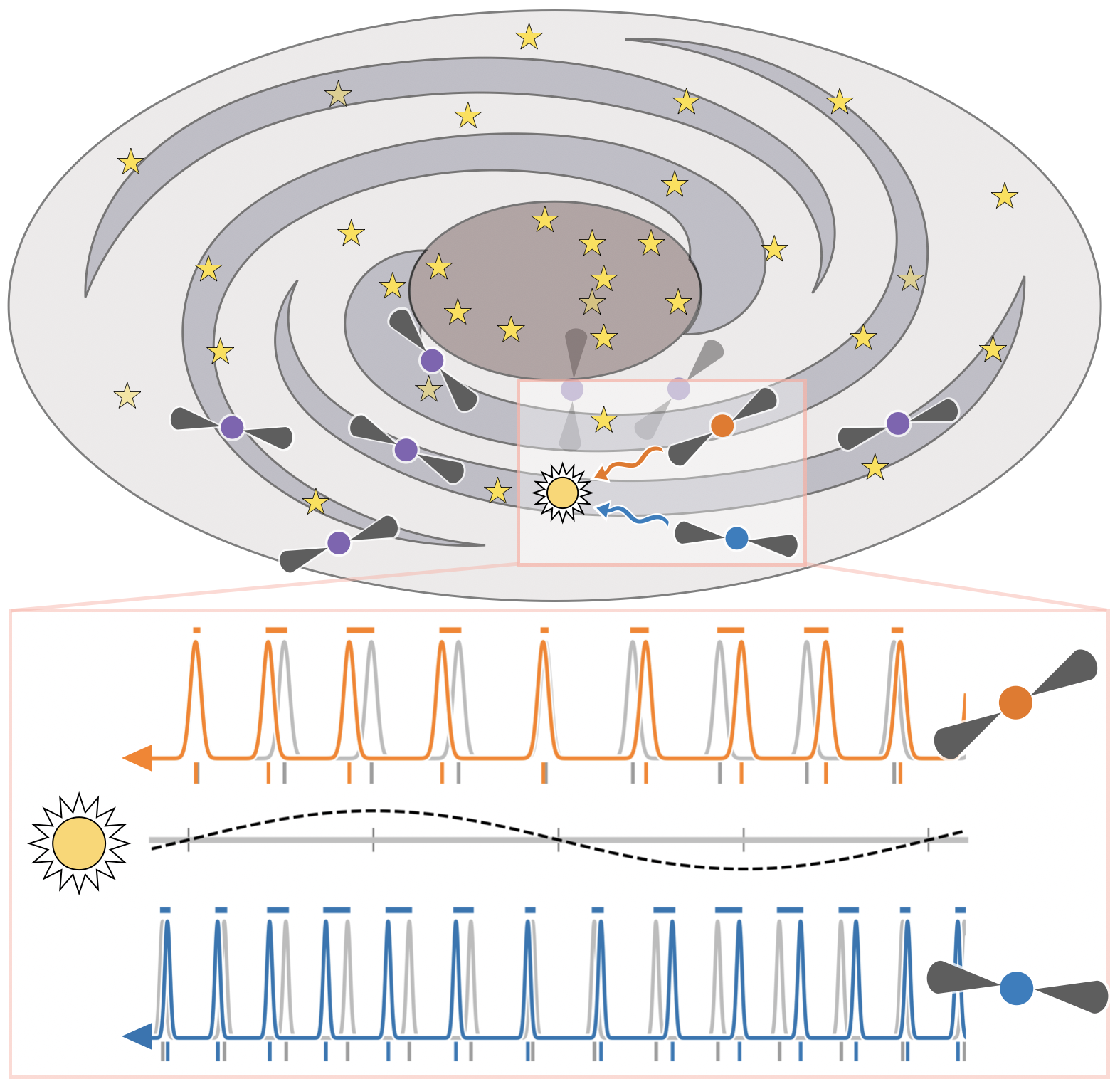}}}
    \caption{Pulsar Timing Arrays are galaxy-scale GW detectors.  Deviations in TOAs are correlated between pulsars to detect GWs.  (Lower:) modeled pulses without GWs (\textit{grey}) vs.~measured pulses with GWs (\textit{blue/orange}); vertical ticks show corresponding TOAs.  Colored horizontal bars are the deviations/residuals proportional to GW amplitude (\textit{black dashed}).}
    \label{fig:galaxy-pta-toas}
    \vspace{-10pt}
\end{wrapfigure}
Instead of lasers, PTAs use \term{pulsars}: spinning neutron stars\footnote{Neutron stars are the collapsed remnants of moderately-massive stars (initial stellar masses of very roughly $\sim 10 - 20 \, \msol$) after undergoing supernovae.  These compact objects have masses of $\approx 2-3 \, \msol$, radii of $\approx 10 \, \tr{km}$, and often enormous magnetic fields ($\approx 10^{12} - 10^{14} \ \tr{Gauss} = 10^8 - 10^{12} \ \tr{Tesla}$).} which produce pulsed bursts of radio emission as their lighthouse-like beams of radiation pass by the observer (Earth) during each rotation.  \term{Millisecond pulsars} in particular, which spin hundreds of times per second, can produce particularly reliable pulses.  Changes in light-travel distance are measured as changes in the \term{times of arrival (TOAs)} of these pulses: either coming earlier or later than they should have in the absence of GWs (Sec.~\ref{sec:pulsar-timing}), shown schematically in the lower-panel of Fig.~\ref{fig:galaxy-pta-toas}.  A wide variety of noise sources can also produce timing deviations (changes in TOAs) that obscure GW signals, but again the specific patterns of timing delays can be correlated between multiple pulsars distributed across the sky/galaxy to distinguish GWs from noise.  The GW frequencies at which PTAs are sensitive is determined by the timing characteristics of pulsar observations.  Data has been collected for $10s~\yr$, so the lowest accessible frequency (the `Rayleigh' frequency) is $f_\tr{lo} \sim \lr[-1]{10~\yr} \sim \nHz$.  The most sensitive frequencies for PTA are near the lowest bin, reaching strain sensitivities on the order of $h \sim 10^{-15}$.  Because pulsar data is unevenly sampled and dominated by noise at high frequencies, the highest sensitive frequency is poorly defined.  In practice, there are currently meaningful constraints up to $f_\tr{hi} \sim 100~\nHz$.

The `smoking gun' for detecting GWs in PTAs is a particular pattern of correlations between different pulsars across the sky: the \term{Hellings \& Downs (HD) correlations} (Sec.~\ref{sec:timing-deviations-spatial-correlations}).  The HD correlation predicts exactly the amount of correlation (or anti-correlation) between the timing delays in two pulsars, based only on their angular separation on the sky.  Pulsars with small angular separations ($\gamma \lesssim 10s \deg$) will be strongly correlated (orange and blue pulsars in Fig.~\ref{fig:galaxy-pta-toas}), those at nearly right angles ($\gamma \sim 90 \deg$) will be anti-correlated, and pairs on opposite sides of the sky ($\gamma \sim 180 \deg$) will again be strongly correlated.

Uncorrelated signals consistent with GWs were first measured by the North American Nanohertz Observatory for Gravitational waves (NANOGrav) PTA in 2019 \citep{Arzoumanian+2020}.  The predicted Hellings \& Downs correlations were then measured with varying confidence by NANOGrav \citep{Agazie+2023}, the Parkes PTA \citep[PPTA;][]{Reardon+2023} centered in Australia, and the European PTA (EPTA) and Indian PTA (InPTA) \citep{Antoniadis+2023}.  The most-recently published confidence levels correspond to a false-alarm probability of $10^{-3} - 10^{-4}$, corresponding to a Gaussian `sigma' level of $3-4$; still below the `$5\sigma$ level' commonly used for a definitive `detection'.  Unlike terrestrial interferometers, PTAs become more sensitive primarily by increasing their data volumes: through the number of pulsars being observed, and the number of TOA measurements from each pulsar\footnote{Higher sensitivity instruments with better noise characteristics, and the discovery of pulsars with low intrinsic-noise can also produce jumps in sensitivity, however significant data must still be collected to probe sufficiently low frequencies.}.  Thus our confidence in whether or not the signal(s) being observed are definitively GWs will gradually improve over the next few years, and particularly through the combination of regional datasets into combined International PTA (IPTA) datasets \citep{Antoniadis+2022}.

The idea of pulsar timing and kindred `doppler tracking' were first formulated in the 1970s and 1980s \citep{Estabrook+Wahlquist-1975, Sazhin-1978, Detweiler-1979, Hellings+Downs-1983}.  At the same time, GWs were first `indirectly' measured by observing the inspiral of a binary pulsar---the Hulse-Taylor pulsar, for the discovery of which the Nobel prize in physics was awarded to Hulse and Taylor in 1993.  Gravitational waves are optimally produced by two, orbiting, massive, point-like objects---like neutron stars or black holes (Sec.~\ref{sec:binary_gws}).  Those GWs carry away energy which has to come from the binary orbit, thus leading the orbit itself to contract until eventual coalescence.  The presence of a pulsar in this double neutron-star system allowed the orbital period to be measured to incredible precision, such that the inspiral rate of $\approx 2.4\E{-12} \, \tr{s/s}$ could be observed \citep{Weisberg+2010}.  GWs from dozens of stellar-mass binaries, containing NSs and black holes with masses of $\sim 10\rm{s} - 100 \, \msol$ have now been \textit{directly} measured by LIGO-Virgo \citep{LVK-2023}.  These kilohertz GWs are produced during the final handful of orbits before these binaries coalesce.

At roughly the same time as pulsar timing was being devised, and GWs indirectly measured, the scientific community was becoming confident in the existence of \term{super-massive black holes (SMBHs)} in the centers of galaxies.  These behemoth objects have masses of $\sim 10^6 - 10^{10} \, \msol$, but event horizon radii of only $10^{-2} - 10^{+2} \, \tr{AU}$, i.e.~only the size of solar systems\footnote{The Astronomical Unit (AU) is the average distance from the Earth to the Sun, with $1 \, \tr{AU} = 1.5\E{11} \, \tr{m} = 4.8\E{-6} \, \pc$.}.  Bright radiation had long been observed across the electromagnetic (EM) spectrum coming from the centers of galaxies: so called \term{active galactic nuclei (AGNs)}, with the brightest versions (quasars) observed from even the distant Universe.  It was quickly realized that one of the only viable power sources for this emission is accretion onto a massive compact object.
% Evidence for such massive compact objects in galaxy centers was observed based on steeply increasing velocity profiles in the centers of galaxies: the speed at which gas and stars orbited near the centers of galaxies required large masses squeezed into those small regions.  Careful measurements of the motion of stars in our own galactic nucleus (Sag $A*$) definitely showed that so much mass was required, in a sufficiently compact region of space, that only a black-hole could exist \needcite{}.  This discovery yielded the 2020 Nobel Prize in Physics to Ghez, \fix{Genzel}, and \fix{Penrose}.  In \fix{2019}, the SMBH in Sgr $A*$ was directly observed by the Event Horizon Telescope, with the SMBH in \fix{Other galaxy} to follow in \fix{2020}.
Careful measurements of the motion of stars in our own galactic nucleus (Sagittarius A-star) demonstrated that only an SMBH could possess such a large mass in such a small space.  This discovery yielded the 2020 Nobel Prize in Physics to Ghez, Genzel, and Penrose.  In 2019, the SMBH in the galaxy M87 was directly observed by the Event Horizon Telescope \citep{EHT-2019}, with the SMBH in our own galaxy, Sagittarius $A*$, following in 2022 \citep{EHT-2022}.

While the existence of SMBHs was still being debated, it was proposed that pairs of SMBHs could form binaries that also produce GWs \citep[e.g.][]{Sazhin-1978, Begelman+1980}.  These \term{SMBH Binaries (SMBHBs)} would be produced following the merger of galaxies, each containing an SMBH (Sec.~\ref{sec:mbhbs}).  Modeling of SMBHB formation and evolution suggests that primarily the most-massive SMBHBs in the Universe, $M \gtrsim 10^9 \, \msol$, are detectable by PTAs.  Kepler's law allows us to relate the binary separation ($a$) to orbital frequency:
\begin{subequations}
\begin{align}\label{eq:kepler_freq}
    f_\tr{orb} & = \frac{1}{2\pi} \scale[1/2]{G M}{a^3}, \\
        & \approx 19 \, \tr{nHz} \, \scale[1/2]{M}{3\E{9} \, \msol} \, \scale[3/2]{a}{10^{-2} \, \pc}.
\end{align}
\end{subequations}
Thus SMBHBs with total masses of $M \sim 10^9 \, \msol$ emit $\nHz$ GWs when separated by $\sim 10^{-2}~\pc$.  In this regime, detectable SMBHBs are still orbiting in the Keplerian regime: only very slowly inspiralling such that they will persist for $\sim 10^6~\yr$ before finally coalescing.  Despite these long lifetimes, the galaxies containing such massive SMBHs are quite uncommon, and they only experience a handful of comparable-mass galaxy mergers in their lifetimes (Sec.~\ref{sec:mbhbs}).  This makes detectable SMBHBs exceedingly rare.  A large number of \term{electromagnetic (EM)} surveys searching for signatures of SMBHBs (Sec.~\ref{sec:ems}) have been carried out; but, despite hundreds of candidates identified to date, not a single SMBHB has been confirmed.

The recent detection of low-frequency GWs by PTAs, appears to be in the form of a \term{stochastic gravitational wave background (GWB)}: the superposition of GW signals from thousands-to-millions of individual SMBHBs.  This GWB provides the first strong evidence that SMBHBs are actually able to form, produce GWs, and eventually coalesce.  However, there are a number of alternative explanations for the GWB that do not involve SMBH binaries at all: signals from so-called `new' or `Beyond the Standard Model (BSM)' physics in the very early Universe\footnote{While mostly outside the scope of this document, we briefly comment discuss non-standard models in Sec.~\ref{sec:cosmo}}.  The binary model is currently strongly favored, but far from proven.  At the very least, PTA observations still provide a new, independent constraint on models that attempt to describe fundamental properties of our Universe such as the nature of cosmic inflation, the origin of the baryon asymmetry (dominance of matter over anti-matter), or provide a quantum description of gravity.  With more data and more sensitivity, PTAs will be able to definitively establish the origin of the GWB, and constrain the origin and evolution of SMBH binaries and/or BSM physics.

Currently, PTAs are able to measure the basic shape of the \term{GWB spectrum}---the amplitude and slope of the GW-strain versus frequency curve.  As PTA sensitivities improve, we will be able to characterize the full shape of the spectrum, including any low-frequency and/or high-frequency deviations from a constant slope, and stochastic variations from one frequency-bin to the next.  We will also be able to observe how the GWB power is distributed over the sky: whether all of the energy is uniformly distributed (isotropic), or coming from some regions more than others (anisotropy), and eventually we will be able to distinguish individual loud binaries, called `\term{continuous wave} (CW)' sources, which are promising targets for EM counterparts.

There are tremendous theoretical uncertainties about nearly every aspect of SMBHB formation and evolution, due to the lack of previous observational constraints.  How are two SMBHs able to find each-other and form binaries in turbulent, post-merger galaxies?  Once large-separation binaries form, how are they able to reach the relatively small separations necessary for detectable GW emission?  Are these binaries able to accrete gas like single SMBHs, and what are the dynamics of that accretion?  If and when accretion occurs, do SMBHBs produce `normal' AGN emission?  What are the EM signatures unique to SMBHBs as opposed to single SMBHs?  How do SMBHBs effect their host galaxies, and what happens following their coalescence?  These are the questions that we hope to answer using PTA observations of low-frequency GWs, and eventually their EM counterparts.

% ---- Gravitational Waves
% ------------------------------------------------------------------------------

\section{Gravitational Waves (GWs)}\label{sec:gws}

\subsection{Waves in General Relativity}\label{sec:gws_gr-waves}

\introref{For additional details, we refer the reader to \citet{MTW-1973}, and the chapter in this volume on GW production (Dror 2025) and detection (Farr 2025).}

The structure and shape of spacetime is described by the \term{metric tensor} $\gmn$, such that the invariant length element of spacetime can be calculated as,
\begin{align}\label{eq:metric}
    ds^2 & = -c^2 d\tau^2 = \gmn \, dx^\mu dx^\nu \equiv \sum_{\nu=0}^{3} \sum_{\mu=0}^{3} \gmn \, dx^\mu dx^\nu = dx_\nu dx^\nu.
\end{align}
The metric is a $4 \times 4$ symmetric tensor where each index runs over the four spacetime dimensions\footnote{The final two equalities above show: (i) Einstein summation notation, where repeated indices imply a summation, and (ii) that metrics are used to `raise' and `lower' indices, transforming between co-variant and contra-variant vector components.}, for example $x_\mu = \lr{c t, x, y, z}$
% {\color{red}JR: I believe $t$ should be $ct$, since you aren't using units where $c=1$.  this may effect other equations!!}
could denote Cartesian coordinates.

We will assume that we are always far from sources of strong gravity, such that we can express our spacetime metric as a perturbation ($\hmn$) to flat spacetime, i.e.~\mbox{$\gmn = \etamn + \hmn$}.  Flat spacetime is typically described by the Minkowski metric $\etamn$, such that {$\etamn \, dx^\mu \, dx^\nu = - c^2 dt^2 + dx^2 + dy^2 + dz^2$}.  We will only consider a \textit{transverse traceless gauge} where $h^\mu_\mu = h_{\mu 0} = 0$, and $\partial_i \hij = 0$.  Einstein's equations determines the relationship between the structure and curvature of spacetime, and its mass/energy content.  The linearized equations in a vacuum admit plane-wave solutions of the form\footnote{It will be assumed that measurables correspond to the real part of complex quantities.},
\begin{align}\label{eq:plane-wave}
    \hmn = A_a  \; \poltens^a_{\mu \nu} \, \exp\!\lr{i \, k_\rho x^{\,\rho}}.
\end{align}
Here, the wave four-vector \mbox{$k_\rho = (2\pi f/c, k_1, k_2, k_3)$} describes the propagation of a GW with frequency $f$ traveling in a spatial direction with components\footnote{We adopt the convention that Greek indices ($\mu, \nu, \dots$) imply the four space-time dimensions, while Latin indices ($i,j,k,\dots$) span only the three space dimensions.} $k_i$.  GWs possess two independent polarizations, here indexed by $a$, described by polarization tensors: $\poltens^a_{\mu \nu}$.  Let us define two unit vectors orthogonal to each other and to the wave-vector (i.e.~GW propagation direction): $\hat{u}_\mu$, $\hat{v}_\mu$, such that \mbox{$\hat{u}_\mu \, \hat{v}^\mu = \hat{u}_\mu \, \hat{k}^\mu = \hat{v}_\mu \, \hat{k}^\mu = 0$}.  We can then define linear polarization tensors as\footnote{The reader may notice that $\hat{u}_\mu$ or $\hat{v}_\mu$ are not fully specified: they can be rotated about $\hat{k}$ by any angle.  This degree of freedom is often specified by the polarization angle, usually denoted by $\psi$.},
\begin{subequations}
\begin{align}
    \eplus_{ij} \equiv & \; \hat{u}_i \hat{u}_j - \hat{v}_i \hat{v}_j, \\
    \ecross_{ij} \equiv & \; \hat{u}_i \hat{v}_j + \hat{v}_i \hat{u}_j,
\end{align}
\end{subequations}
each corresponding to an amplitude $A_a \in \{A_+, A_\times\}$.
% Any GW plane-wave can be described as a combination of two linearly polarized components ($\eplus, \ecross$ with amplitudes $A_+, A_\times$ or $h^+, h^\times$) or two circularly polarized components ($\eright, \eleft$, with amplitudes $A_R, A_L$).
These waves follow null geodesics, such that $k_\mu \, k^\mu = \etamn \, k^\mu \, k^\nu = 0$, and thus $(2\pi f/c)^2 = k_i k^i$.  In cartesian coordinates with the GW traveling along the $\hat{z}$ direction (i.e.~\mbox{$k_\mu = \{2\pi f/c, 0, 0, k_z\}$}), we can write,
\begin{align}\label{eq:gw-metric-simple}
    \hmn(t,z) =
        \begin{pmatrix}
        0 & 0 & 0 & 0 \\
        0 & A_+ & A_\times & 0 \\
        0 & A_\times & -A_+ & 0 \\
        0 & 0 & 0 & 0
        \end{pmatrix} \, \exp\lrs{2\pi i f \lr{t - z/c}}.
\end{align}
The effects of each polarization are `\term{quadrupolar}', i.e.~repeating every $180 \deg$, with the $\times$ polarization equal to the $+$ rotated by $45 \deg$.  The effects of the two polarizations on a ring of test-masses are shown in Fig.~\ref{fig:binary-pulsar-gw}(c).

% However, it is often necessary to work in externally imposed coordinates.  Often we will describe the source position in spherical coordinates, which we can orient with $\hat{u}_i$ in the polar direction and $\hat{v}_i$ in the azimuthal direction,
% \begin{subequations}
% \begin{align}\label{eq:wave-basis}
%     \hat{r} \equiv - \hat{k} & = \sinpar{\theta} \cospar{\phi} \, \hat{x} + \sinpar{\theta} \sinpar{\phi} \, \hat{y} + \cospar{\phi} \, \hat{z}, \\
%     \hat{\theta} = \hat{u} & = \cospar{\theta} \cospar{\phi} \, \hat{x} + \cospar{\theta} \sinpar{\phi} \, \hat{y} - \sinpar{\phi} \, \hat{z}, \\
%     \hat{\phi} = \hat{v} & = -\sinpar{\phi} \, \hat{x} + \cospar{\phi} \, \hat{y}. \\
% \end{align}
% \end{subequations}
% See [MTW1973 Sec 35.5]; [Maggiore-2007 Sec 1.3]
In general relativity, the `geodesic equation' governs motion due to gravity: \mbox{$\odv[2]{x^\alpha}{\tau} + \Gamma^\alpha_{\mu \nu} \odv{x^\mu}{\tau} \odv{x^\nu}{\tau} = 0$}, where $\Gamma^\alpha_{\mu \nu}$ are `Christoffel symbols'---functions of the metric and its derivatives.  Given a GW metric, in the transverse traceless gauge and linearized gravity, all of the Christoffel symbols vanish, and thus the geodesic equations become \mbox{$\odv[2]{x^\alpha}{\tau} = 0$}.  This tells us that the \textit{coordinate} motion of test masses remain constant before, during, and after GWs pass.  The same is not true for the \textit{proper distance} between two test masses.  From the metric (Eq.~\ref{eq:metric}), we can write for any spacetime interval:
\begin{align}\label{eq:spacetime-interval-integration}
    s = \int_{x(t)} \lrs[1/2]{\gmn \odv{x^\mu}{t} \odv{x^\nu}{t}} dt.
\end{align}
Consider test-mass A to be at coordinate position $x^A_i = 0$, and test-mass B to be at a coordinate position $x^B_i = \distcoord \hat{x}_1$.  If we reuse the coordinate system with a GW traveling along the $\hat{x}_3$ direction (Eq.~\ref{eq:gw-metric-simple}), and recall that proper distance $L$ is the spacetime interval defined with $dx^0 = 0$, we can find that:
\begin{align}\label{eq:length-simple}
    L = \int_0^{\distcoord} \lrs[1/2]{1 + h_{11}(t)} dx^1 \approx \lr{1 + \frac{1}{2} \hplus \, \exp\!\lrs{2\pi i f \lr{t - z/c}} }~\distcoord.
\end{align}
The approximation comes from a first-order Taylor expansion of the integrand.  If we consider the `strain'---the fractional proper-distance change---over a full wave cycle, we find \mbox{$h \equiv \Delta L / L = \hplus$}.  We can then identify the GW metric perturbation ($h_{ij}$) as a direct proxy for the local, fractional deformation, or the \term{gravitational wave strain}.  The local deformations are thus quadrupolar, just like the waves themselves.

It becomes clear that the proper-distance is what's relevant for experiments\footnote{Light-travel measurements determine not only delay times (e.g.~pulsar/Doppler tracking) and equivalently phase shifts (e.g.~laser interferometers), but also dynamical forces such as beads on a string or equivalently masses on a spring (e.g.~resonant detectors).} if we consider that we could identically calculate the light-travel time in A's reference frame, for a null geodesic traveling from A to B, by replacing the left-hand side of Eq.~\ref{eq:spacetime-interval-integration} with $c \Delta T$.  We can thus equivalently identify GW strain as \mbox{$h \equiv \Delta T / T$}.

\subsection{Timing Deviations and Spatial Correlations}\label{sec:timing-deviations-spatial-correlations}

\introref{For additional details, we refer the reader to \citet{Anholm+2009} and \citet{Romano+Allen-2024}.  Below, we closely follow the latter.}

\begin{wrapfigure}{r}{0.6\textwidth}
    \vspace{-6pt}
    \includegraphics[width=0.6\textwidth]{{{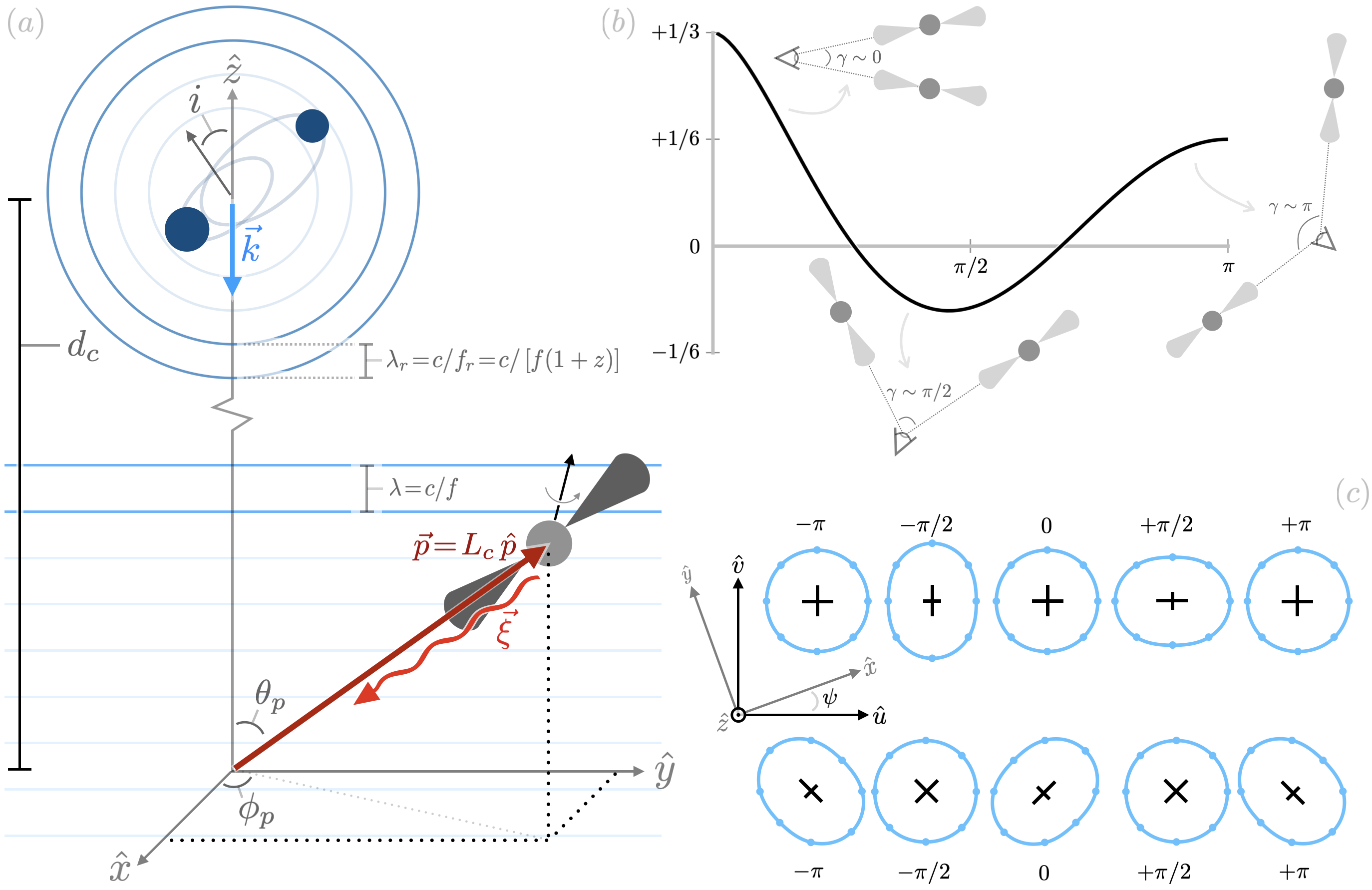}}}
    \caption{(a) Binary at comoving distance $d_c$ along the $\hat{z}$ axis, emitting GWs along the $\vec{k}$ vector; and pulsar at position $\vec{p}$ emitting pulses along the path $\vec{\xi}$.  (b) `Hellings-Downs Curve': signal correlation between pulsars vs.~angle between them, $\hdfunc(\gamma)$.  (c) Two polarization patterns vs.~GW phase from $\left[-\pi, +\pi\right]$, for `plus $+$' (upper row) and `cross $\times$' (lower row) polarizations.}
    \label{fig:binary-pulsar-gw}
    \vspace{-2pt}
\end{wrapfigure}

The preceding calculation (Eq.~\ref{eq:length-simple}) requires the GW to be spatially uniform along the direction of interest (i.e.~$h_{ij}[x^1,t] = h_{ij}[x^1\!=\!0,t]$).  In general this is not the case, and both the amplitude and frequency of the GW could be changing.  Consider a pulsar located at a spatial position $p^i$, which is a total coordinate distance from the observer $\distcoord = \lr{p_i \, p^i}^{1/2}$.  The fully general solution, using light travel time, is:
% \citep{Romano+Allen-2024} Eq. 13a; [Rakhmanov-2009]
\begin{align}\label{eq:gw-timedelay-integral}
    c \, \Delta T(t) = \frac{1}{2} \hat{p}^i \hat{p}^j \int_{0}^{\distcoord} \hij\lrs{\xi\lr{s}} \, ds,
\end{align}
 where the metric is integrated along a path specified by $\xi(s)$.  The configuration is shown in Fig.~\ref{fig:binary-pulsar-gw}(a).  We will see below that the same GW effects can be considered more conveniently with respect to a `redshift':
% \footnote{In terms of a change in frequency, this would equivalently be $z(t) = \lr{\nu(t) - \nu_0}/\nu_0$.}
\begin{align}\label{eq:gw-redshift-integral}
    z(t) \equiv \odv{\Delta T}{t} = \frac{1}{2} \frac{\hat{p}^i \hat{p}^j}{c} \int_{0}^{\distcoord} \pdv{\hij\lrs{\xi\lr{s}}}{t} \, ds.
\end{align}
Without a loss of generality, the path that light takes from the pulsar to the observer can be parameterized as, \mbox{$\xi(s) = \lr{t - \frac{1}{c}\lrs{\distcoord - s}, \lrs{\distcoord-s}\hat{p}^i}$}.
For a plane wave, we can re-parameterize in terms of `retarded coordinates', $h\lrs{\xi\lr{s}} = h\lrs{u\lr{s}}$, where
\begin{align}
    u\lr{s} \equiv t\lr{s} - \hat{k}^i \, \xi_i\lr{s} / c = t - \lr{\distcoord - s}\lr{1 + \hat{k}^i \, \hat{p}_i} / c.
\end{align}
It is easy to evaluate Eq.~\ref{eq:gw-redshift-integral} by expressing $\partial \hij / \partial t$ in terms of $\partial \hij / \partial s$; i.e.~we want to find the coordinate-time rate of change of the metric strain in terms of the rate of change along the trajectory.  If we assume that the frequency is constant along the trajectory, this is straightforward because $\hij$ only depends explicitly on the `retarded coordinate' $u$, which depends linearly on $t$ and $s$.  Thus, we can write:
\begin{subequations}
\begin{align}
    \pdv{h}{s} & = \odv{h}{u} \pdv{u}{s} = \frac{1}{c}\lr{1 + \hat{k}^i \hat{p}_i}, \\
    \pdv{h}{t} & = \odv{h}{u} \pdv{u}{t} = \odv{h}{u} = \frac{c}{\lr{1 + \hat{k}^i \hat{p}_i}} \pdv{h}{s}.
\end{align}
\end{subequations}
Finally, plugging this into Eq.~\ref{eq:gw-redshift-integral} we get:
\begin{align}\label{eq:gw-redshift} % \citep{Romano+Allen-2024} Eq. 18
    z(t) = \frac{1}{2} \frac{\hat{p}^i \hat{p}^j}{1 + \hat{k}^i \hat{p}_i} \lrs{\hij(t, 0) - \hij(t - \distcoord/c, p)}.
\end{align}
Thus, the redshift is simply the difference of metric strains between the pulsar (the \textit{pulsar term}) and the observer (the \textit{Earth term})\footnote{Note that, even for constant frequency, this is not the case for time-delays: the expression in Eq.~\ref{eq:gw-timedelay-integral} cannot be reduced in general as there is an additional build-up of time/phase change between the endpoints of the trajectory.  Under the long-wavelength approximation, $\distcoord \ll k^{-1} = \lambda$, the integrand can be taken as constant, and we obtain a projected version of Eq.~\ref{eq:length-simple} which is appropriate for high-frequency, ground-based GW interferometers.}.

Equation~\ref{eq:gw-redshift} describes the response of timing measurements to a passing GW plane-wave, with constant frequency.  Note the factor of $1 + \hat{k}^i \hat{p}_i$ which appears in the denominator, and also in the `phase' of the GW strain.   The situations in which the pulsar is aligned or anti-aligned with the source, $\hat{p} = \pm \hat{k}$, are worth particular consideration.  When the pulsar is in the direction that the GW is coming from ($\hat{p} = -\hat{k}$), the redshift goes to zero despite divergence in the denominator, because the numerator is also zero.  In a way, the numerator is zero for two separate reasons.  First, there is no component of the GW strain along the observer-pulsar direction for this configuration.  Second, the propagating photons are always `surfing' the same phase of the GW \citep{Anholm+2009}, so it necessarily arrives at the observer at the same GW phase as it left the pulsar.  However, in the case that the source and pulsar are anti-aligned, $\hat{p} = +\hat{k}$, the phase of the GW changes as the photons propagate to the observer, and thus the metrics differ between the Earth and Pulsar\footnote{Note, however, that the overall response is still zero, but in this case only for 'one reason'---that the strain is again zero along the observer-pulsar direction.}.  This nicely illustrates that the phase information is not symmetric to flipping the GW propagation direction.  While the GWs are purely quadrupolar, the response of a detector is \textit{not purely quadrupolar}.  See \citet{Romano+Allen-2024} for an excellent discussion.

It will be convenient to rearrange Eq.~\ref{eq:gw-redshift} by grouping the phase and amplitude terms such that $h^a_{ij} = h^a \poltens^a_{ij}$, and also introducing \textit{geometrical projection factors} (or \textit{antenna response functions}) $F^a(k_i)$:
\begin{subequations}
\begin{align}\label{eq:gw-redshift-projection}
    z(t) & = \sum_a F^a(k_i, p_i) \cdot \lr{h^a_E - h^a_P}, \\
    F^a(k_i, p_i) & \equiv \frac{1}{2} \frac{\hat{p}^i \hat{p}^j}{1 + \hat{k}^i \hat{p}_i} \poltens^a_{ij}, \\
    h^a & = A_a \, \exp\lr{i \, k_\alpha x^\alpha} = A_a \exp\lrs{2\pi i~\lr{f t - k_i x^i}}.
\end{align}
\end{subequations}
Here we have substituted $h^a_E = h^a(t, 0)$ for the Earth term in each polarization, and $h^a_P = h^a(t - \distcoord/c, p)$ for the pulsar term.
% and also using \textit{detector response functions} $R^a(k^\mu, p^i)$,
% \begin{align}\label{eq:gw-redshift-response}
%     z(t) = e^{-2\pi i f t} A_a R^a \lr{h^a(E) - h^a(P)}, \\
%     R^a(k^\mu, p^i) \equiv \lrs{1 - e^{-2\pi i f L \lr{1 + \hat{k}^i \hat{p}_i}}} F^a(k_i, p_i),
% \end{align}
% where $A_a$ are again the two polarization amplitudes $A_+,A_\times$.
The amplitude of a GW signal embedded in pulsar timing data will typically be far smaller than the amplitude of noise (Sec.~\ref{sec:noise-sources}).  The \textit{measured} redshift for a pulsar $p$, including noise $n_p(t)$ for that pulsar, is:
\begin{align}\label{eq:redshift_with_noise}
    z_p(t) = \sum_a F^a_p \, h^a_E - F^a_p \, h^a_P + n_p(t).
\end{align}
The key to pulsar timing \textit{arrays} is examining the cross-correlations between a pair of pulsars $p$ and $s$ (Sec.~\ref{sec:os}):
\begin{align}\label{eq:redshift-cross-correlation}
    \rho_{ps}(\tau) = \left< z_p(t) \, z_s(t - \tau) \right>_T \equiv \frac{1}{\tobs} \int_{-\tobs/2}^{\tobs/2} z_p(t) \, z_s(t - \tau) \, dt,
\end{align}
specifically at zero time-lag, $\rho_{ps} \equiv \rho_{ps}(\tau=0)$.  Here, we are integrating over a total observing-duration $\tobs \sim 10 \, \yr$.  Because the number of GW wave-cycles between the Earth and each pulsar (and also between different pulsars) is large, $f L / c \gg 1$, this means that the pulsar term and the Earth term will be effectively independent, and the pulsar terms from different pulsars will also be nearly independent\footnote{For individual GW plane-waves, the pulsar terms will be correlated at time-lags corresponding to their light-travel distances (for the Earth terms) or the differences in light-travel times (between pulsar-terms).  Including the pulsar term(s) can thus double the amount of signal when searching for individual sources, but this is outside of the scope for this discussion.}.  If the noise is also independent between different pulsars, then,
\begin{align}\label{eq:redshift-correlation-approx}
    \rho_{ps} \approx \sum_a \frac{F^a_p \, F^a_s}{T} \int_{-T/2}^{T/2} \big| h_{E} \big|^2 dt \approx P_h \sum_a F^a_p \, F^a_s.
\end{align}
Here, $P_h$ is approximately the (average) signal-`power' of $h(t)$ in the data-stream, which becomes exact in the limit that $\tobs \rightarrow \infty$.

It is convenient to consider the Fourier transform of signals, and their cross-correlations.  The Fourier transform is,
\begin{align}\label{eq:fourier-transform}
    \tilde{h}(f) \equiv \lim_{\tobs\rightarrow \infty} \int_{-\tobs/2}^{+\tobs/2} h(t) \, e^{-2 \pi i f t} \, d t.
\end{align}
Here we use $\tilde{h}(f)$ to emphasize that this is a Fourier transform, however we will later follow the convention used in the literature that both the time-domain and frequency-domain GW-strain use the same symbol $h: h(t) \leftrightarrow h(f)$, which is motivated by sinusoidal waveforms.  We can use Parseval's theorem to relate the signal-power in time and frequency, and thereby define the power spectral density $S_h(f)$:
\begin{align}\label{eq:psd_fourier}
    P_h & = \frac{1}{\tobs} \int_{-\tobs/2}^{+\tobs/2} \left| h(t) \right|^2 dt = \frac{1}{\tobs} \int_{-\infty}^{+\infty} \left| \tilde{h}(f) \right|^2 df \equiv 2 \int_{0}^{+\infty} S_h(f) \, df.
\end{align}
For a stationary process, i.e.~one whose statistical properties are independent of time, each frequency is uncorrelated such that,
\begin{align}
    % \left< \tilde{h}(f) \, \tilde{h}(f') \right> = S_h(f) \, \delta(f - f'),
    \left< \tilde{h}(f) \, \tilde{h}(f') \right> =
    \begin{cases}
        S_h(f) & f = f', \\
        0      & f \neq f'.
    \end{cases}
\end{align}
This allows us to define frequency-specific cross-correlations as,
\begin{align}\label{eq:redshift-correlation-frequency}
    \rho_{ps}(f) \approx S_h(f) \sum_a F^a_p \, F^a_s.
\end{align}

\citet{Hellings+Downs-1983} showed that the cross-correlation could be averaged over a large number of unpolarized GW sources, distributed uniformly across the sky, to obtain\footnote{We designate the average over the sky as $\langle \dots \rangle_\Omega$ and the average over time as $\langle \dots \rangle_T$.}:
\begin{subequations}\label{eq:hd}
\begin{align}
    \langle \rho_{ps} \rangle_\Omega & = \langle \big| h_E \big|^2 \rangle_T \, \hdfunc(\gamma) = P_h \, \hdfunc(\gamma), \\
    \langle \rho_{ps}(f) \, \rangle_\Omega & = S_h(f) \, \hdfunc(\gamma), \\
    \hdfunc(\gamma) & \equiv \frac{1}{3} - \frac{1}{6} \scale{1 - \cospar{\gamma}}{2} + \scale{1 - \cospar{\gamma}}{2} \ln \scale{1 - \cospar{\gamma}}{2}. \label{eq:hd_hd}
\end{align}
\end{subequations}
The \term{Hellings-Downs correlation}, $\hdfunc(\gamma)$, depends entirely on the angle between pulsars: $\cospar{\gamma} = \hat{p}_{p,i} \, \hat{p}_s^i$.  The Hellings-Downs curve, shown in Fig.~\ref{fig:binary-pulsar-gw}(b), is the `smoking gun' for evidence of a GW signal in PTA data.

For a comprehensive discussion of HD correlations, see \citet{Romano+Allen-2024}, which we have largely followed.  A few additional points are worth noting.  \citet{Cornish+Sesana-2013} show that Eqs.~\ref{eq:hd} also holds for a single GW source, averaging over a large number of uniformly distributed pulsars\footnote{c.f.~\citet{Romano+Allen-2024} however, who show that averaging over pulsars can be more nuanced than averaging over sources.}, i.e.~$\langle \rho_{ps} \rangle_\Omega = \langle \rho_{ps} \rangle_{ps}$.  \citet{Allen-2023} shows that the standard deviation in the HD correlations can be comparable to the mean, even in the limit of large numbers of sources; and derive expressions for the mean and variance of the correlations with and without polarization, and with and without the pulsar terms.

\subsection{GWs from Binaries}\label{sec:binary_gws}

\noindent\textit{For additional details, we refer the reader to \citet[][Part~VIII]{MTW-1973}, \citet[][Ch.~4]{Flanagan+Hughes-1998}, and \citet{Enoki+Nagashima-2007}.}

\subsubsection{Single Binary GW Power and Evolution}\label{sec:gw-power-evolution}

In general, GW production must be calculated numerically.  However, an analytic solution exists in the limits of (i) `slow motion'\footnote{For gravitating systems this is equivalent to the `quadrupole approximation': that the size-scale of the objects and their motion is small compared to the emitted wavelength.}, i.e.~$v \ll c$; and (ii) that the gravitating objects are moving pseudo-periodically.  A full derivation is an involved process, but the results for a binary in nearly-Keplerian orbit\footnote{`Nearly-Keplerian' in the sense that the orbital elements change on timescales much longer than the binary orbital period.} are surprisingly simple \citep{Peters+Mathews-1963}.  The power radiated, or luminosity $L$, into a solid angle $\Omega$ from each polarization $a$, can be expressed in terms of the third time derivative of the quadrupole moment tensor ($Q_{ij}$),
\begin{subequations}
\begin{align}\label{eq:power-quadrupole}
    \frac{dL^a}{d\Omega} & = \frac{G}{8\pi c^5} \lrs[2]{\frac{d^3 Q_{ij}}{dt^3} \poltens^a_{ij}}, \\
    Q_{ij}(t) & \equiv \int \lrs{x_i \, x_j - \frac{1}{3} \delta_{ij} x_k \, x^k} \, \rho(t,x) \, d^3 x,
\end{align}
\end{subequations}
where $\rho$ is the mass density.  For a binary in a circular orbit, GWs are emitted at twice the orbital frequency.  In this case, the rest-frame total luminosity, summed over both polarizations and integrated over all angles, is:
\begin{subequations}\label{eq:lum_gw_circ}
\begin{align}
    \lgwcirc & = \frac{32}{5 G c^5} \left(G\mchirp\right)^{10/3} \left( 2\pi \frstorb \right)^{10/3}, \\
        & \approx 8.4\E{39} \, \mathrm{W} \, \scale[10/3]{\mchirp}{10^9 \, \msol} \, \scale[10/3]{\frstorb}{3~\mathrm{nHz}}
\end{align}
\end{subequations}
where $\frstorb$ is the rest-frame orbital frequency.  Much of GW emission is determined by a particular combination of the two component masses $m_1$ and $m_2$, called the `\term{chirp mass}',
\begin{align}
    \label{eq:chirp_mass}
    \mchirp = \frac{\left(m_1 m_2\right)^{3/5}}{M^{1/5}} = M \frac{q^{3/5}}{\left(1 + q\right)^{6/5}},
\end{align}
where the total mass is $M = m_1 + m_2$, and the mass ratio is $q = m_2/m_1 \leq 1$  We define all masses to be in the rest frame.  From Eqs.~\ref{eq:lum_gw_circ}, we see that the amount of energy radiated by GWs is enormous: comparable to the EM energy radiated from a large galaxy or bright quasar.  However, the dynamical effects of these GWs, and thus their detectability, is quite small due to the intrinsically weak coupling of the gravitational force.

Binaries with an eccentricity $e$ emit GWs at all integer harmonics $n$ of the orbital frequency.  This allows us to decompose the luminosity into contributions from each harmonic $n$,
\begin{subequations}
\begin{align}
    \lumgw & = \sum_{n=1}^\infty L_n = \sum_{n=1}^\infty \lgwcirc \, g(n,e) = \lgwcirc \, F(e), \\
    g(n,e) & \equiv \frac{n^4}{32} \Biggl(
        \begin{aligned}[t]
            & \left[ J_{n-2}(ne) - 2eJ_{n-1}(ne) + \frac{2}{n} J_n (ne) + 2e J_{n+1} (ne) - J_{n+2} (ne) \vphantom{\frac{2}{n}}\right]^2 \\
            & + \left(1-e^2\right) \Bigl[J_{n-2}(ne) - 2eJ_n(ne) + J_{n+2}(ne)\Bigr]^2 + \frac{4}{3n^2}\Bigl[J_n(ne)\Bigr]^2 \Biggr),
        \end{aligned} \\
    F(e) & \equiv \frac{1 + \frac{73}{24} e^2 + \frac{37}{96} e^4}{\left( 1 - e^2\right)^{7/2}}.
\end{align}
\end{subequations}
Here, $F(e)$ is the `eccentricity function', $g(n,e)$ is the `frequency distribution function', and $J_n(x)$ is the $n$th-order Bessel function of the first kind.  Note that for zero eccentricity: $g(n\!=\!2,e\!=\!0) = 1$ and $g(n\!\neq\!2, e\!=\!0) = 0$.

During the psuedo-Keplerian phase, the energy emitted from a binary as gravitational waves comes from the orbital energy\footnote{During the `chirp' and final coalescence of the binary, this is no longer entirely true, and up to $\sim 5\%$ of the rest-mass energy of the binary can be emitted as GWs.}.  This allows us to calculate the `hardening'\footnote{The term `hardening' is adopted from the stellar binary literature where `soft' binaries at larger separations tend to separate further due to three body interactions, while `hard' binaries at smaller separations tend to come closer together: the so-called `Heggie' or `Heggie-Hills' law.} rate, at which the binary inspirals.  Let us define the `\term{hardening timescale}' (or `\term{residence timescale}') with respect to frequency as: $\tau_f \equiv dt/d\ln(f) = f/\lr{df/dt} = -\frac{2}{3} a/\lr{da/dt}$.  For gravitational wave emission \citep{Peters-1964},
\begin{subequations}\label{eq:hard_time_gw_freq}
\begin{align}
    \taufgw \equiv \frac{f}{\lrs{df/dt}_\tr{gw}} & = \frac{5}{96} \scale[-5/3]{G \mchirp}{c^3} \frac{\lr{2 \pi \frstorb}^{-8/3}}{F(e)}, \\
        & \approx 4.5\E{5} \, \yr \, \scale[-5/3]{\mchirp}{10^9 \, \msol} \, \scale[-8/3]{\frstorb}{3 \, \tr{nHz}} \, \lrs[-1]{F(e)}.
\end{align}
\end{subequations}
Note that the total binary lifetime to go from $\frstorb$ to coalescence ($f\rightarrow \infty$) is notably smaller: $\approx (3/8)~\taufgw$.  The same chirp-mass at a relatively high frequency of $\approx 30 \, \tr{nHz}$ would still have a residence time of $970 \, \yr$.  Thus, for typical binary parameters, PTA sources can be considered as monochromatic over human lifetimes\footnote{Because the light-travel time from a pulsar can be $\sim 10^3 \, \yr$, the pulsar may see a noticeably different GW frequency than the Earth.  Thus inclusion of the `pulsar term' (i.e.~Eq.~\ref{eq:gw-redshift}) can be used to measure the `chirp', and thus directly constrain the SMBHB mass, independently of its distance.}.  When discussing SMBH binary evolution (Sec.~\ref{sec:binary_evolution}), it is convenient to write the total binary lifetime in terms of binary semi-major axis: $\tau_\tr{life} = \int_{a_0}^{a_1} \lr[-1]{da/dt} da$.  Due to GW emission, this is $\approx (a/4)/(da/dt)$ or exactly,
\begin{subequations}\label{eq:lifetime_gw_sepa}
\begin{align}
    \taulifegw & = \frac{5 \, c^3}{256 \, G^3} \scale{a_0^4 - a_1^4}{M \, m_1 \, m_2} \\
        & \approx 2.3~\tr{Gyr} \, \scale[5/3]{\mchirp}{10^9~\msol} \scale[4/3]{M}{3\E{9} \, \msol} \scale[4]{a}{0.4~\pc}.
\end{align}
\end{subequations}
Due to the steep dependence on separation ($a^4$), the `final' separation $a_1$ can be taken as zero (c.f.~Sec.~\ref{sec:coalescence}).  GWs also carry away angular momentum, which can be related to the binary eccentricity.  The eccentricity-decay timescale is,
\begin{align}\label{eq:eccen_time}
    \tauegw \equiv \frac{e}{-\lrs{de/dt}_\tr{gw}} = \frac{15~c^5}{304~G^3} \frac{a^4}{M~m_1~m_2} \lr[-1]{1 + \frac{121}{304} e^2}
\end{align}
which is very similar to the lifetime in terms of separation.

It will be convenient to know the spectrum of GW energy emitted by a binary.  We will denote the rest-frame GW frequency of the binary as it evolves over time as $f_r'(t)$, and use a delta-function to pick-out the time at which the binary is emitting at the rest-frame frequency of interest $f_r$:
\begin{subequations}\label{eq:binary_gw_energy_spectrum}
\begin{align}
    \frac{d\energygw(f_r)}{d \ln\!f_r} & = \int dt~\frac{d^2 \energygw}{dt~d\ln\!f_r'}~\delta\lrs{f_r'(t) - f_r}, \\
        & = \sum_{n=1}^{\infty} \lgwcirc(\frstorb) \, \frac{dt}{d\ln\!f_r} \frac{g(n,e)}{n} \Bigg|_{\frstorb = f_r/n}.
\end{align}
\end{subequations}
The second equality requires the binary to \textit{reach} the orbit corresponding to the GW frequency of interest.  This is not the case when the orbital frequency is: (i) below the frequency at which the binary becomes gravitationally bound (see Sec.~\ref{sec:mbhbs}), (ii) above the frequency at which the binary coalesces\footnote{Eccentric binaries emit at higher frequencies, but SMBHBs are virtually guaranteed to be circular by the time of coalescence.  After coalescence, some GWs can still be emitted during the `ring down', but the total energy emitted in this phase is small.}, or otherwise (iii) that the binary evolution stalls before reaching that frequency.  Given those caveats, we can calculate the GW energy spectrum for a circular binary, assuming purely GW-driven binary evolution (i.e.~$dt/d\!\ln\!f_r = \taufgw$) to find,
\begin{align}\label{eq:binary_gw_energy_spectrum_gw_circ}
    \frac{d\energygw(f_r)}{d \ln\!f_r} \Bigg|_{\tr{gw}} & = \frac{\lr[5/3]{G \mchirp}}{3~G} \lr[2/3]{2 \pi f_r} \Bigg|_{f_r = 2 \, \frstorb}.
\end{align}

\subsubsection{GW Strain}\label{sec:gws_strain}

The GW strain can be calculated directly from the second time-derivative of the quadrupole moment tensor as,
\begin{align}\label{eq:strain-quadrupole}
    \hij(t) = \frac{2 G}{c^4 \, \comdist} \frac{\partial^2 Q_{ij}(t - \comdist/c)}{\partial t^2},
\end{align}
where we have noted explicitly that the quadrupole-moment tensor must be evaluated at the `retarded time' when the GW was emitted.  Here $\comdist$ is the `comoving distance' to the source.  The GW power can be related to the strain of each harmonic as,
\begin{align}
    \label{eq:strain_lum}
    \hsn^2(\fgw) = \frac{G}{c^3} \, \scale[2]{2}{n} \, \frac{L_n}{\lr[2]{2 \pi \, \frstorb} \, \comdist^2}
            = \hscirc^2 \, g(n,e) \, \scale[2]{2}{n}.
\end{align}
In Eq.~\ref{eq:strain_lum} we have shifted to describing the strain as a function of frequency instead of time, with the two related by the Fourier transform (Eq.~\ref{eq:fourier-transform}).  Note also that the observer-frame GW frequency is $\fgw = n \, \frstorb / (1 + z)$, for a source at a cosmological redshift $z$.  The average strain for a binary in a circular orbit can be expressed as,
\begin{align}
    \label{eq:strain_circ}
    \hscirc(\fgw) & = \frac{8}{10^{1/2}}\, \frac{\left(G\mchirp\right)^{5/3}}{c^4 \, \comdist} \lr[2/3]{2\pi\frstorb}, \\
        & \approx 7.9\E{-16} \, \scale[5/3]{\mchirp}{10^9 \, \msol} \scale[2/3]{\frstorb}{3 \, \tr{nHz}}.
\end{align}
This `source strain', defined for each harmonic, is a scalar value: the strain averaged over both polarizations and over all solid angles (c.f.~the strain tensor of Eq.~\ref{eq:plane-wave} or Eq.~\ref{eq:strain-quadrupole}).  The time-dependent strain of each polarization at all viewing angles and non-zero eccentricity can be found in \citet{Barack+Cutler-2004}.

When considering the detectability of GW signals, it is common to utilize the GW `\term{characteristic strain}' $h_c$ instead of the raw GW strain $h$ or $h_s$ \citep[for thorough discussions, see:][]{Flanagan+Hughes-1998, Moore+2015}.  The goal is to construct a measure of GW strength ($h_c$) that will be proportional to the square-root of the \term{signal-to-noise ratio (SNR)} for a given detector.  The square-root ensures that the GW energy is still $\propto h_c^2$.  This measure of strain is ill-defined instantaneously: it is intrinsically \textit{integrated}, over some observing time and possibly also frequency bandwidth.  For nearly-monochromatic sources, the appropriate quantity is,
\begin{align}\label{eq:characterist_strain_mono}
    h_c^2(f) = f \, \tobs \, h_s^2(f),
\end{align}
where the power of the signal is increased by the number of cycles over an observing duration $\tobs$.  This is usually the quantity most relevant for PTAs.  The characteristic strain for frequency-evolving sources is instead,
\begin{align}\label{eq:characterist_strain_chirp}
    h_{c,\tr{chirp}}^2(f) = \frac{2 f^2}{df/dt} h_s^2(f),
\end{align}
where the power is increased by the number of cycles near the frequency $f$.  More generally, the characteristic strain can be related to the (one-sided) power spectral density ($S_h$) as, $h_c^2 = f \, S_h(f)$.  This is sometimes taken as the definition of characteristic strain.  In practice, the power spectral density would be calculated by measuring the time-dependent strain $h(t)$, and taking the Fourier transform to find $\tilde{h}(f)$, which can then be used to estimate $S_h(f)$ using Eq.~\ref{eq:psd_fourier}.  In this way, the power spectral density and characteristic strain are still well defined for stochastic signals (discussed below).  Particularly in cosmological contexts, it is common to describe signals in terms of the GW energy density ($\energydensgw$) per logarithmic frequency interval,
\begin{align}
    \Omega_\tr{gw}(f) \equiv \frac{1}{\rho_c \, c^2} \frac{d \energydensgw}{d \ln f} = \frac{2 \pi^2}{3 H_0^2} f^3 S_h(f).
\end{align}
Here, the GW-energy spectrum\footnote{The terminology is confusing here.  Note that the `power' spectral density is a generalized power of a (dimensionless) signal in data, with units of inverse frequency.  The GW-energy density, $\mathcal{E}_\tr{gw}$ is a \textit{physical}-energy per unit volume, and $\rho_c$ is a mass-density: mass per unit volume.} is normalized by the cosmological critical density, $\rho_c = 3 H_0^2 / 8 \pi G$, for a redshift-zero Hubble constant $H_0$.

\subsubsection{The GW Background from a Population of Binaries}\label{sec:gws_gwb}

As shown in Eqs.~\ref{eq:hard_time_gw_freq}, the lifetime of SMBH binaries in the PTA band is long ($\gtrsim 10^5~\yr$), and we might expect a number of binaries to be emitting in each frequency bin (discussed further in Sec.~\ref{sec:mbhbs}).  The combination of a large number of individual sources with different frequencies and phases leads to a `\term{stochastic GW Background (GWB)}'.  \citet{Phinney-2001} presented an elegant method to calculate the local/present-day energy density of GWs based on the integrated history of GW emission from binaries over cosmic time\footnote{As \citet{Phinney-2001} points out, this is a GW-version of the `\term{Soltan argument}': if quasar luminosity is accretion powered, then the integrated energy in quasar light over the history of the Universe, divided by the fraction of rest-mass energy radiated away, must match the present day mass-density of SMBHs (the remnants of past quasars).}.  Specifically, we can write that,
\begin{align}\label{eq:phinney_argument}
    \frac{d \energydensgw(f)}{d \ln f} = \frac{\pi~c^2}{4~G} \, f^2~h_c^2(f) = \int_0^\infty \frac{dz}{1+z} \frac{d^2 n}{d\mchirp~d z} \, \frac{d \energygw(\mchirp, f_r)}{d \ln f_r} \Bigg|_{f_r = f(1+z)}.
\end{align}
The comoving number density of binaries is $n \equiv dN/dV_c$, where $V_c(z)$ is the comoving volume at a redshift $z$, and the $1+z$ term in Eq.~\ref{eq:phinney_argument} accounts for the redshifting of GW energy from emission to observation.  Here $\energydensgw{}$ is the local energy per unit comoving volume of GWs, while $E_\tr{gw}$ is the total energy emitted in GWs from a particular binary, and its spectrum is given by Eqs.~\ref{eq:binary_gw_energy_spectrum}.  Assuming GW-only evolution of a smooth continuum of circular binaries (Eq.~\ref{eq:binary_gw_energy_spectrum_gw_circ}) we find,
\begin{subequations}\label{eq:gwb_idealized}
\begin{align}
    h_c^2(f) & = \frac{4\pi}{3c^2} \lr[-4/3]{\pi f} \int \int dz \, d\mchirp \, \frac{d^2 n}{d\mchirp~dz} \, \frac{\lr[5/3]{G\mchirp}}{\lr[1/3]{1+z}}, \\
    h_c(f) & \approx 1.1\E{-15} \, \scale[-2/3]{\fgw}{1\,\pyr} \, \left( \frac{ n_\tr{eff}}{10^{-4}~\tr{Mpc}^3} \scale[5/3]{\mchirp_\tr{eff}}{10^9~\msol} ~\lr[-1/3]{1+z}_\tr{eff} \right)^{1/2}. \label{eq:phinney_gwb_spectrum}
\end{align}
\end{subequations}
The second relation includes the insight that each merger leads to a redshift-zero remnant SMBH, so that the local number-density (or mass-density) of SMBHs can be directly tied to the GWB amplitude.  The second relation in Eq.~\ref{eq:phinney_gwb_spectrum} relies on \textit{ad hoc} effective/integrated parameter values.  We have neglected terms that describe the fraction of local mass built-up by mergers.  A precise and insightful treatment can be found in \citet{Sato-Polito+2024}.

\begin{figure}
    \vspace{-10pt}
    \includegraphics[width=\textwidth]{{{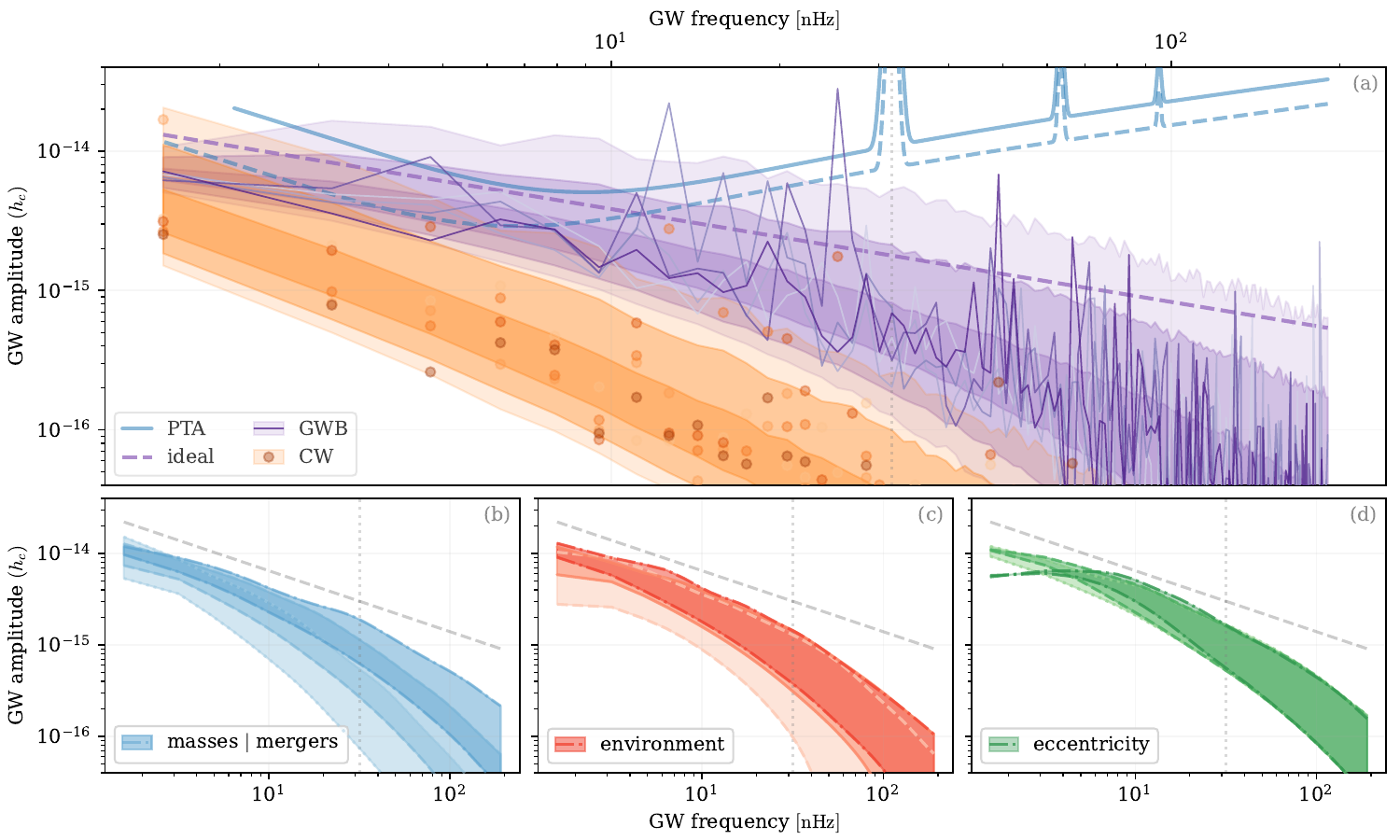}}}
    \caption{\textbf{(a) GWs from binary populations.} GWB (\textit{blue}): with $50\%, 90\%, 98\%$ intervals (contours), and five random realizations (\textit{lines}); loudest CW in each frequency bin (\textit{orange}): same contours and random realizations (\textit{circles}).  Idealized GWB spectrum (\textit{purple dashed}) from Eq.~\ref{eq:phinney_gwb_spectrum}.  Schematic PTA sensitivity curves (blue) for 15yrs (solid) and 20yrs (dashed) of data.  \textbf{(b) Effects of SMBH masses and merger rates} on GWB (blues), showing $50\%$ intervals for three population models, each calibrated to the same idealized amplitude - decreasing merger rates as masses increase.  \textbf{(c) Effects of environmental hardening} on GWB (reds), $50\%$ intervals for models with three different strengths of environmental interactions at sub-parsec separations.  \textbf{(d) Effects of binary eccentricity} on GWB (greens), $50\%$ intervals of varying binary eccentricities ($0.5, 0.9, 0.95$ initialized at $0.1 \, \pc$).}
    \label{fig:gwb}
    \vspace{-10pt}
\end{figure}

The most important feature of the GWB from SMBH binaries is a characteristic-strain spectrum $h_c(f) \propto f^{-2/3}$.  Often the spectrum is characterized/fit as a power-law, normalized at a frequency of $1~\pyr \approx 31~\tr{nHz}$, i.e.~$h_c(f) \approx A_{\pyr} \scale[-2/3]{f}{1~\pyr}$.  Idealized, power-law GWB spectra are plotted as dashed lines in Fig.~\ref{fig:gwb}; purple in panel (a) and grey in panels (b),(c),(d).  Realistic models of SMBH binary populations show considerable deviations from this idealized power-law.  GWB spectra from five random realizations of binary populations are shown as purple lines in Fig.~\ref{fig:gwb}(a), while $50\%, 90\%, \& 98\%$ confidence intervals are shown with purple contours.

There are four primary causes of deviation from the idealized power-law of Eq.~\ref{eq:gwb_idealized}:
\begin{enumerate}
    \item The GWB is formed by a discrete number of binaries, with variations in the number of binaries from bin-to-bin (`cosmic variance', `Poisson variations', or `shot noise'), which will produce deviations from a uniform spectral index.  This is shown in Fig.~\ref{fig:gwb}(a).
    \item At sufficiently high frequencies (typically $f \gtrsim 30 \tr{nHz}$), the expected number of binaries contributing to the GWB will drop below unity, and the spectrum will steepen.  Higher-mass binaries will also coalesce, and no longer contribute GWs at the highest frequencies.  Fig.~\ref{fig:gwb}(b) shows the effects on the spectrum of changing the relative number of binaries vs.~their masses, contributing to the GWB.
    \item At low frequencies (large separations), binary evolution is not entirely GW-driven, with noticeable contributions from `environmental' processes (Sec.~\ref{sec:binary_evolution}).  This increases the rate of binary inspiral, which decreases the total GW deposition, and thus attenuates/decreases the GWB spectrum at low frequencies (see Eq.~\ref{eq:binary_gw_energy_spectrum}).  The effectiveness of environmental interactions also determines the fraction of binaries able to reach the PTA band.  Varying environmental interaction strengths are shown in Fig.~\ref{fig:gwb}(c).
    \item Eccentric binaries distribute their emitted GW energy across a range of frequencies, instead of only at twice the orbital frequency.  In populations that have considerable eccentricity in the PTA band, this decreases the GW energy at low frequencies, and increases the GW energy at high frequencies.  At the same time, eccentric binaries generally inspiral more quickly than circular binaries.  These effects are expected to be relatively small.  Fig.~\ref{fig:gwb}(c) compares the GWB spectrum from different eccentricity binary populations.
\end{enumerate}
Each type of deviation, when measured by PTAs, provides significant additional information about the underlying SMBHB population.  Thus it is largely the \textit{deviations} from the idealized power-law behavior which can provide the most information from low-frequency GW observations\footnote{While outside the scope of this article, yet more information is encoded in the occurrence rate of individually detectable `CW' binary sources, and the closely-related anisotropy of GWB emission across the sky.}.

These effects can be taken into account by: first, recasting Eq.~\ref{eq:phinney_argument} in terms of the number of binaries at each frequency ($dN/d\ln f_r$), instead of their number-density \citep{Sesana+2008, Sesana-2013}; and second, by self-consistently evolving binaries from formation (large separations) to the frequencies of interest (smaller separations) while including binary eccentricity \citep{Kelley+2017a, Kelley+2017b}.  The latter point will be discussed further in Sec.~\ref{sec:binary_evolution}.  The key argument to the former is explicitly connecting each binary's evolution over frequencies to the time evolution of the Universe:
\begin{subequations}
\begin{align}
    \diffp{N}{{\mchirp}{z}{\ln f_r}} & = \diffp{N}{{\mchirp}{z}{V_c}} \diffp{{V_c}}{{z}} \diffp{{z}}{{t_r}} \diffp{t_r}{{\ln f_r}} \\
        & = \diffp{n}{{\mchirp}{z}} \, 4\pi c \, d_c^2 \, \lr{1+z} \, \tau_f.  \label{eq:number_density_to_number_frequency_b}
\end{align}
\end{subequations}
Plugging this expression into Eq.~\ref{eq:phinney_argument}, along with the circular and GW-only approximations, gives us the result that,
\begin{align}\label{eq:gwb-number-quadrature}
    \left< h_c^2(f) \right> = \int d\mchirp~dz~\diffp{N}{{\mchirp}{z}{\ln f_r}} \, h_s^2(\mchirp, z, f_r)\Bigg|_{f_r = f (1+z) / 2}.
\end{align}
In other words, the GWB characteristic strain is given by the quadrature sum of the strains from all binaries per logarithmic frequency interval.  We have added angle brackets on the left-hand side of Eq.~\ref{eq:gwb-number-quadrature} to emphasize that this yields an expectation value, or average GWB strain spectrum.  This comes both from extrapolating a number-density of binaries to a total number in the Universe (Eq.~\ref{eq:number_density_to_number_frequency_b}), and also from adding GW strains in quadrature which neglects interference between individual binaries and different phases.

\subsubsection{Coalescence Rates and Memory Effects}\label{sec:coalescence}

During the final few orbits of binary inspiral, the `slow motion' approximation becomes increasingly inaccurate and the GW emission relations presented in Sec.~\ref{sec:gw-power-evolution} break down.  This happens near the mutual `innermost stable circular orbit (ISCO)' of the binary.  The ISCO location for SMBH binaries, is often approximated as that of an extreme mass-ratio inspiral with zero spin(s):
\begin{align}
    \aisco \equiv \frac{6~GM}{c^2} \approx 8.6\E{-4} \, \pc \, \scale{M}{3\E{9}~\msol} \hspace{40pt}
    \fisco \approx 730 \, \nHz \, \scale[-1]{M}{3\E{9}~\msol}.
\end{align}
While, technically, such high frequencies could be probed by PTAs, in practice the sensitivity is entirely swamped by noise and the detection of a chirp is unfeasible.  Additionally, the coalescence rate of such high-mass binaries is expected to be vanishingly small (Sec.~\ref{sec:mbhbs}).  The coalescence rate can be calculated in a similar manner to the number of binaries (Eq.~\ref{eq:number_density_to_number_frequency_b}),
\begin{subequations}
\begin{align}\label{eq:gwrate}
    \diffp{N}{{t}} & =
            \int dz \, \diffp{N}{{z}{V_c}} \diffp{V_c}{{z}} \diffp{z}{{t_r}} \diffp{t_r}{{t}} \\
        & = \int dz \, \diffp{n}{{z}} \, 4 \pi \, c \, d_c^2.
\end{align}
\end{subequations}
When such coalescences occur, they produce not only a GW chirp, but also a shift or `DC offset' in the space-time metric which returns to a different rest state than it started \citep{Thorne-1992}.  This produces a broad-band signal called a GW `\term{burst with memory}'.  The characteristic amplitude can be estimated as (\textit{Ibid.}),
\begin{align}
    h_{c,\trt{BWM}} \approx 10^{-17} \, \scale{\epsilon_\trt{BWM}}{10^{-2}} \, \scale{\mchirp}{10^9~\msol} \, \scale{\comdist}{500~\tr{Mpc}},
\end{align}
where $\epsilon_\trt{BWM}$ is an efficiency parameter that depends on orientation and BH spins.  For more information, see, for example, \citet{Favata-2010}.

% ---- Pulsar Timing Arrays
% ------------------------------------------------------------------------------

\section{Detection of Gravitational Waves with Pulsar Timing Arrays}\label{sec:ptas}

Before considering the detection of GWs, consider the much simpler case of placing limits on the amount of GW power at the Earth, using pulsar measurements.  We can limit the local GW spectral strain based on the fractional, maximum deviation in pulse \term{times-of-arrival (TOAs)}.  We will see below that (i) we will be highly noise dominated, and (ii) that we will be interested in the superposition of many GWs which can interfere with each other.  For these reasons, instead of relying on a single maximum deviation, we measure the mean-squared deviations which are proportional to the signal power (e.g., Eq.~\ref{eq:psd_fourier}), and then limit the GW power to the same amount:
\begin{align}
    S_h \propto \langle \Delta h^2 \rangle = \sigma_h^2 \lesssim \frac{\langle \Delta t^2 \rangle}{\tobs^2} = \frac{\sigma_{\Delta t}^2}{\tobs^2} \propto S_\trt{TOA}.
\end{align}
Here, $\Delta t$ is the TOA deviation, $\tobs$ is the total observing duration, and $\sigma_h$ and $\sigma_{\Delta t}$ are the standard-deviations in strain and TOA deviations respectively\footnote{The variance of a quantity $x$ is, \mbox{$\sigma_x^2 \equiv \langle x^2 \rangle - \langle x \rangle^2$}, and both strain and timing residuals have zero means s.t.~$\sigma_x^2 \approx \langle x^2 \rangle$.}

Typically we will characterize GW power in terms of a power spectrum over frequencies $S_h(f)$ (Sec.~\ref{sec:timing-deviations-spatial-correlations}).  Recall that the measurable GW frequencies are limited to falling between the Rayleigh and Nyquist frequencies, i.e.~$\tobs^{-1} \lesssim f \lesssim \lr{2\,\dtobs}^{-1}$, for a typical time between TOA measurements (`sampling interval' or `cadence'), $\dtobs$.  Specifically, the Fourier frequency basis is $f_i = i / T$, for integer values $i \in \{1, \dots, N_f\}$.  The minimum frequency, and the frequency bin-width, is the `Rayleigh' frequency, $f_1 = \tobs^{-1}$.  The maximum frequency for evenly sampled data is the `Nyquist' frequency, $f_{N_f} = \lr{2\,\dtobs}^{-1}$, such that the number of independent frequencies is $N_f \approx T / \lr{2\, \dtobs}$, although $f_{N_f}$ becomes poorly defined for irregularly sampled data, which is typically the case for PTAs.

\subsection{Pulsar Timing}\label{sec:pulsar-timing}

\noindent\textit{For additional details, we refer the reader to \citet{Romano+Cornish-2017} and \citet{Taylor-2021}.}

The accuracy by which TOAs can be measured is determined by a combination of the intrinsic pulse stability (determined by largely-unknown pulsar physics) and the radiometer measurement accuracy, i.e.~$\sigmatoa^2 \sim \sigma_p^2 + \sigmaradio^2$.  For general information on pulsar astronomy, we direct the reader to \citet{Lorimer+Kramer-2012}.  Empirically, the best millisecond pulsars can have stabilities of $\sim 100~\ns$.  The measurement accuracy can be estimated as,
\begin{align}
    \scale[2]{\sigmaradio}{w} \approx \scale{w}{P - w} \, \scale[-1]{S_\tr{pulsar}}{S_\tr{radio}} \, N_\tr{radio}^{-1},
\end{align}
i.e.~the precision of the pulsar TOA ($\sigmaradio$) over the pulse width ($w$), is the ratio of the pulse width to non-pulse (`noise') for pulsar period $P$, divided by the ratio of power from the pulsar signal to the instrument noise, divided by the number of measurements.  This is called the `radiometer equation'.  The number of measurements (samples) can be expressed as a product of observing duration, bandwidth (i.e.~data sampling rate), and number of independent polarizations (two): \mbox{$N_\tr{radio} = T_\tr{TOA} \cdot \Delta f \cdot 2$}.  We can then rewrite the radiometer equation adopting typical values as\footnote{The instrument noise is often expressed as a product of a `system temperature' and `instrument gain': $(T_\tr{radio} / 50~\mathrm{K}) = (S_\tr{radio} / 10~\mathrm{J}) \cdot (G_\tr{radio} / 5~\mathrm{K~Jy}^{-1})$.  In radio astronomy the `Jansky' unit of spectral flux density is sadly common, \mbox{$1~\mathrm{Jy} \equiv 10^{-23} \textrm{ erg s}^{-1} \textrm{ cm}^{-2} \textrm{ Hz}^{-1} = 10^{-26} \textrm{ J s}^{-1} \textrm{ m}^{-2} \textrm{ Hz}^{-1}$}.}:
\begin{align}\label{eq:radiometer-equation-fiducial}
    \sigmaradio \approx 100~\ns \; \scale[3/2]{w}{1~\ms} \scale[-1/2]{S_\tr{pulsar}}{1~\mathrm{mJy}} \scale[1/2]{S_\tr{radio}}{10~\mathrm{Jy}} \scale[-1/2]{T_\tr{TOA}}{30 \, \min} \scale[-1/2]{P}{10~\ms} \scale[-1/2]{\Delta f}{100~\mathrm{Mhz}}.
\end{align}
For millisecond pulsars, the single-burst signal-to-noise ratio is often on the order of $S_\tr{pulsar}/S_\tr{radio} \sim 10^{-4}$, and thus large numbers of pulses must be stacked or `folded' to become detectable.  Choosing an observing duration of $\sim 30 \, \min$ gives us an instrumental precision matching the intrinsic stability of millisecond pulsars.  The optimal PTA `sensitivity' to gravitational waves over a decade of observing time is then $h_\trt{PTA} \approx \langle \Delta h^2 \rangle^{1/2} \approx 10^{-16} - 10^{-15}$.  For a careful analysis of PTA sensitivities, see \citet[][which includes sensitivity forecasting]{Siemens+2013} and \citet[][which includes the construction of sensitivity curves]{Hazboun+2019}.

It is convenient to express pulse TOAs in terms of different components: a deterministic component (which may include any GW signals of interest) and a stochastic `noise' component\footnote{Particular effects are often grouped as either `noise' or `deterministic' depending on the particular formalism.  Ultimately the requirements is that the noise components can be described by a noise covariance matrix (described below).},
\begin{align}\label{eq:toa-components}
    \dt(t) = \overline{\Dt}(t) + n(t).
\end{align}
The goal of pulsar timing is to capture the deterministic component through a \term{timing model}, $\Dt(t|\alpha) \approx \overline{\Dt}(t)$, with model parameters $\alpha$.  The `\term{timing residuals}', the differences between observations and the timing model, are then:
\begin{align}\label{eq:residual-timing-model}
    r\lr{t|\alpha} \equiv \dt(t) - \Dt\lr{t|\alpha}.
\end{align}
The timing model is fit to the data by constructing a likelihood that assumes the residuals to be distributed as multidimensional Gaussian noise:
\begin{align}\label{eq:toa-likelihood}
    p\lr{r|\Dt} = \lrs[-1/2]{\det\lr{2\pi \, N_{ij}}} \exp \lrs{ -\frac{1}{2} r_i \, N^{-1}_{ij} \, r_j }.
\end{align}
Because TOAs are evaluated at discrete times $t_i$, we adopt the notation that $r_i \equiv r(t_i)$.  Here $N_{ij}$ is a `noise covariance matrix'\footnote{The term `noise correlation matrix' is often also used in the literature.  The difference is only in their normalizations: $\tr{corr}(x,y) = \tr{cov}(x,y) / \sigma_x \sigma_y$, but for likelihood calculations this has no effect, so the two types of matrices can be used interchangeably.}:
\begin{align}\label{eq:correlation}
    N_{ij} = \tr{cov}(n_i, n_j) \equiv \Big \langle \bigl [ n_i - \langle n_i \rangle \bigr ] \bigl [ n_j - \langle n_j \rangle \bigr ] \Big \rangle
        = \langle n_i \, n_j \rangle - \langle n_i \rangle \, \langle n_j \rangle.
\end{align}
In practice, the timing model is typically first calculated for each individual pulsar alone, ignoring any GW components because $h(t) \ll \langle \Dt \rangle / T$.  This can be done with `maximum likelihood' estimates determined by maximizing Eq.~\ref{eq:toa-likelihood}, or equivalently the `log likelihood', $\ln\lrs{p(r|\Dt)}$.  Alternatively, full probability distributions (`posteriors') for timing parameters can be determined using Bayes' theorem, in this context:
\begin{align}\label{eq:bayes-timing-model}
    p\big(\alpha|\dt,\model\big) = \frac{ p\big(\dt|\alpha,\model\big) \, p\big(\alpha|\model\big) }{p\big(\dt|\model\big)},
\end{align}
where we have made explicit that this requires a model $\model$.  One significant benefit of the latter approach is that uncertainties in the timing model parameters can be marginalized over, at times analytically.  In the Bayesian framework, multiple possible models can be compared with the `odds ratio',
\begin{align}\label{eq:odds-ratio}
    \mathcal{O}_{\mu \nu}\big(\dt\big) \equiv \frac{p\big(\model_\mu \, | \dt\big)}{p\big(\model_\nu \, | \dt\big)}
        = \frac{p\big(\dt \, | \model_\mu\big)}{p\big(\dt | \model_\nu\big)} \frac{p\big(\model_\mu\big)}{p\big(\model_\nu\big)}.
\end{align}
This provides a quantitative means of determining what components should be included in a given timing model.  Both the maximum likelihood and Bayesian methods yield a timing model, and can give us a first estimate of the timing model parameters: $\alpha^\tr{noise}$.

The GW signals we search for are small, so we can generally assume that the timing model parameters in the presence of GWs, $\alpha^\tr{GW}$, can be found with a slight perturbation ($\delta \alpha$) from their values assuming noise alone:
\begin{subequations}\label{eq:linear-timing-model}
\begin{align}
    \Dt\lr{\alpha^\tr{GW}} \approx \Dt\lr{\alpha^\tr{noise}} + M \delta\alpha, \\
    M \equiv \frac{\partial \Dt}{\partial \alpha}\bigg|_{\alpha = \alpha^\tr{GW}}, \;\;\;
    \delta \alpha \equiv \alpha^\tr{GW} - \alpha^\tr{noise}.
\end{align}
\end{subequations}
Here $M$ is usually expressed as an $N_\trt{TOA} \times N_\trt{par}$ matrix for each pulsar, called the `design matrix', which gives the dependency of each TOA ($t_i; i \in [0, N_\trt{TOA}-1]$) on each timing-model parameter ($\alpha_j: j \in [0, N_\tr{par}-1]$), i.e.~$M_{ij} \equiv \partial \Dt_i / \partial \alpha_j$.  This is the `linearized' timing model which is drastically faster to optimize and can often be performed (semi-)analytically.

\subsection{Noise Sources}\label{sec:noise-sources}

\introref{For additional details, we refer the reader to \citet{Backer+Hellings-1986}, \citet{Cordes+Shannon-2010} and \citet[][Ch.~6]{Condon+Ransom-2016}.}

Constructing a noise model, and performing pulsar timing more broadly, is actually quite complicated.  Individual pulses are usually far too faint to detect (Eq.~\ref{eq:radiometer-equation-fiducial}), so large numbers need to be stacked to make a measurement.  Intrinsically, pulses are typically far from delta functions, with substantial substructure.  That substructure can vary over time, sometimes stochastically, which is sometimes referred to as `jitter', sometimes with systematic trends, and sometimes abruptly.  Even a delta-function pulse is subject to the orbit of the pulsar if it is in a binary (which is common), changes in its spin, and propagation effects through the ionized `inter-stellar medium (ISM)'---particularly `dispersion' and `scintillation' which are both strongly radio-frequency dependent.

Typically, the roughly-dozen different physical components of noise can be treated as independent, additive components, such that we can write:
\begin{align}\label{sec:timing-model-components}
    \Dt = \Delta t_\tr{DM} + \Delta t_\tr{sc} + \Delta t_\tr{spin} + \Delta t_\tr{bary} + \Delta t_\tr{rel} + \dots.
\end{align}
These timing model components can be incorporated either in the time-domain or in the frequency/Fourier domain.  Timing components in the frequency basis can still be part of the `linearized' timing-model because the \textit{coefficients} are the model parameters of interest:
\begin{align}
    \Delta t^\tr{freq}_j = F^\tr{freq}_{jk} \alpha^\tr{freq}_k = \sum_{k=0}^{N_f} \alpha^\tr{freq}_k \exp\lr{-2\pi i \ k \ t_j / \tobs},
\end{align}
where $F^\tr{freq}_{jk}$ is the timing model for Fourier-basis frequency $k$ at time $t_j$ and $\alpha^\tr{freq}_k$ is the Fourier coefficient for frequency $k$.  Many of the timing-model components are periodic, and most of the noise components are stationary (i.e.~depending not on individual times, but on time lags: $|t_i - t_j|$), making the Fourier basis particularly convenient.  For a much more thorough description of the construction of a timing model including time-domain and frequency-domain components, see \citet{Taylor-2021}.  A subset of important components of the timing model are briefly described here.
\begin{itemize}
    \item \textbf{ISM dispersion}: a plasma propagation effect which delays EM waves in a EM-frequency ($\nu$) dependent manner \citep{Draine-2011}:
        \begin{subequations}\label{eq:dm-delay}
        \begin{align}
            & \Delta t_\trt{DM}(\nu) \approx \frac{e^2}{2\pi m_e c} \frac{D_m}{\nu^2} \approx 4.6 \, \sec \; \scale{D_m}{100 \, \tr{pc cm}^{-3}} \, \scale[-2]{\nu}{300 \, \MHz} \\
            & D_m \equiv \int_0^L n_e(l) \; dl .
        \end{align}
        \end{subequations}
        The `Dispersion Measure (DM; $D_m$)' is the column-density of free electrons from the observer to the source.  To constrain the DM, measurements must be taken at multiple radio frequencies to fit the $\propto \nu^{-2}$ delays.  The measured arrival times can then be converted to some reference frequency, often taken as $\nu \rightarrow \infty$.  DM is particularly challenging to model as it has substantial spatial and temporal structure that can be common across many different pulsars due to the solar wind.
    \item \textbf{Scintillation / scattering}: Scintillation or scattering is the accumulation of phase changes (or path changes) due to the interaction of radio pulsses with localized over-densities in the ISM (often modeled as clumpy `screens').  The frequency dependence of scintillation depends on the density spectrum of the scattering material, but is typically steeper than dispersion with a characteristic power-law index of $\approx 4$.  For example, for a Kolmogorov spectrum\footnote{i.e.~a PSD with a size-scale power-law index of $-5/3$ for energy, or $-11/3$ for density.} \citep{Rickett-1990}:
        \begin{align}
            \Delta t_\trt{sc}(\nu) \approx 10^{-6}~\sec \, \scale[6/5]{C}{10^{-17} \, \cm^{-20/3}} \scale[-22/5]{\nu}{300 \, \MHz} \scale[11/5]{D}{1 \, \kpc}.
        \end{align}
        The normalization ($C$) and slope of actual density power-spectrum must be measured from observations and can vary significantly.  Additional effects depend on the relative velocities of the observer, source, and scattering screen and thus vary systematically-in-time with additional structure.
    \item \textbf{Pulsar evolution}: pulsars tend to spin down due to loss of spin energy, including from dipole-like emission from their large magnetic fields.  Spin changes are typically parameterized in terms of time-derivatives of the spin period, $\dot{P}, \ddot{P}$, etc.  For magnetic dipole radiation, the $\dot{P}$ effect on timing is,
        \begin{align}
            \Delta t_{spin} \approx \dot{P} \cdot \tobs \approx 1.9\E{-5} \, \sec \, \scale[2]{B}{10^{12}~\tr{G}} \scale{R}{10~\tr{km}} \scale{\tobs}{10\,\yr} \scale[-1]{M_p}{2 \, \msol} \, \scale[-1]{P}{10 \, \ms},
        \end{align}
        where $B, R, M_p, P$ are the pulsar magnetic field, radius, mass, and spin period, respectively.  Additionally, astrometric motion on the sky, and binary motion\footnote{Most millisecond pulsars are also in binaries (or higher-order multiples) as their short spin-periods generally require \textit{spinning up} through mass-transfer from a binary companion.} must also be accounted for.
    \item \textbf{Solar-system motion}: The motion of the Earth (observatory) contributes significantly to TOA measurements.  To make comparisons across time (decades) and space (different observatories on Earth, and orbital positions), pulse TOAs are typically moved to the `solar-system barycenter' reference frame which can be taken as an inertial frame over long periods of time.  This requires accounting for Earth's position ($\approx 500 \, \sec$; the `Roemer delay') and motion ($\approx 10^{-6} \, \sec$) but also that of the other planets using measured ephemerides \citep{Vallisneri+2020}.  For example, if the uncertainty in Jupiter's orbit is $\delta x_\tr{Jup} \approx 100 \, \tr{km} \approx 10^{-3} \, R_\tr{Jup}$, then the added uncertainty in the solar-system barycenter position's light-travel time is,
    \begin{align}
        \Delta t_\tr{bary} \approx 3.2\E{-7} \, \sec \, \scale{M_\tr{Jup}}{\msol} \, \scale{\delta x_\tr{Jup}}{100 \, \tr{km}}.
    \end{align}
    \item \textbf{Relativistic effects}: As pulses pass near massive objects (e.g.~binary companions or our Sun), their propagation is affected by the `Shapiro Delay': the added propagation time across curved space-time.  For a source at an angle $\theta$ to a massive object, the Shapiro delay is \citep{Backer+Hellings-1986}:
        \begin{align}
            \Delta t_\tr{rel} \approx - \frac{2GM}{c^3}\lr{1 - \cos\theta} \approx - 9.8\E{-6} \, \s \, \scale{M}{\msol} \, \lr{1 - \cos\theta}.
        \end{align}
        Additional relativistic effects can also be important.  For example, for very small angular separations, the increased light-travel distance due to gravitational lensing can have a larger effect than the Shapiro delay; and motion of the pulsar and Earth in their respective gravitational potentials changes their gravitational redshifts over time, producing `Einstein delays'.
\end{itemize}

An epoch of observations for a single pulsar will typically be tens of minutes long, collecting $\sim 10^4 - 10^5$ pulses, measuring the DM, scattering, and other measurable noise processes and then constructing a single average/effective pulse arrival time $t_i$ along with a measured uncertainty $\sigma_i$.  After subtracting the timing model (Eq.~\ref{eq:residual-timing-model}), we are left with any GW signals in addition to unmodeled noise---both intrinsic noise and residuals due to imperfect fitting of the deterministic components.  The noise typically has a component that is `white', having uniform power with respect to signal-frequency $f$, and also significant `red'-noise with more power at lower frequencies.  Different pulsars are observed to have very different red-noise characteristics, often referred to as `spin noise', which is typically modeled as a power-law spectrum.

\subsection{The Optimal Statistic \& GW Searches}\label{sec:os}

\introref{For additional details, we refer the reader to \citet{Anholm+2009}, \citet{Romano+Cornish-2017} and \citet{Taylor-2021}.}

We saw in Sec.~\ref{sec:gws} that a distinct feature of GWs is the correlations between detector data-streams.  For PTAs, these are the Hellings-Downs correlations (Eqs.~\ref{eq:hd}; Fig.~\ref{fig:binary-pulsar-gw}) which depend only on the angular separation between pairs of pulsars on the sky.  It is then natural to utilize these correlations as a method, or as a `statistic' to detect them.  In this context, a \term{detection statistic} is a quantity that can be calculated from the measured data, which can be directly related to our confidence in a signal being present---for example a signal-to-noise ratio (SNR) or Bayes factor.

Let us express our pulsar TOAs as including a GW signal component $s(t)$, a timing model that approximates deterministic physical effects $\Dt(t) \approx \overline{\Dt}(t)$ (Eq.~\ref{sec:timing-model-components}), and a noise component $n(t)$:
\begin{align}\label{eq:toa-components-gw}
    \dt(t) & = \overline{\Dt}(t) + s(t) + n(t), \\
    r(t) & \equiv \dt - \overline{\Dt}(t) \approx s(t) + n(t).
\end{align}
Consider two different data streams of residuals, $r_i(t)$ and $r_j(t)$, from two different pulsars.  We can construct the cross-correlation (Eq.~\ref{eq:redshift-cross-correlation}), and consider its expectation value at zero lag\footnote{As mentioned previously: for CW signals, the pulsar terms (Eq.~\ref{eq:gw-redshift}) can be recovered by considering time-lags matching the light-travel times to each pulsar.}:
\begin{align}\label{eq:cross-correlation}
    % C_{ij}(\tau) & \equiv \frac{1}{T} \int_{-T/2}^{+T/2} r_i(t) \, r_j(t-\tau) \, d t, \\
    % C_{ij}(t) & \equiv r_i(t) \, r_j(t), \\
    \langle \rho_{ij} \rangle & = \langle s^2 \rangle + \langle s \, n_i \rangle + \langle s \, n_j \rangle + \langle n_i \, n_j \rangle \approx \langle s^2 \rangle.
\end{align}
The approximation is valid as long as any non-GW correlations between two pulsars have correctly been subtracted in the timing models, and thus the strains and both noise streams are uncorrelated\footnote{Noises and strains are un-correlated both in time and ensembles: $\lim_{T \rightarrow \infty} \rho_{ij} = \langle s^2 \rangle_T$, and also \mbox{$\langle \rho_{ij} \rangle = \left< s^2 \right>_E$}.  However, $\langle s^2 \rangle_T \neq \langle s^2 \rangle_E$, because our `ensemble' is composed of different population realizations.  Variances will also differ in the two cases.}: \mbox{$\langle s \, n_i \rangle \approx \langle s \, n_j \rangle \approx \langle n_i \, n_j \rangle \approx 0$}.  If our data streams are pulsar redshifts, then we can identify Eq.~\ref{eq:cross-correlation} with Eqs.~\ref{eq:hd}, i.e.~\mbox{$\langle \rho_{ij} \rangle_{i \neq j} = \langle \big| h_E \big|^2 \rangle_T \, \hdfunc(\gamma) = P_h \, \hdfunc(\gamma)$}.  The cross-correlation gives us the GW power, filtered by the detector response: the Hellings-Downs correlations.

To make estimates from data, it is convenient to construct a likelihood function.  Let us again consider two, discretely sampled data streams $r_{1i} \equiv r_1(t_i)$, and $r_{2i} \equiv r_2(t_i)$ each with $\nsamp$ samples\footnote{In practice, some sort of interpolation or binning must be performed to get samples at matching times, unless we are working in frequency-space (see below).}, which we concatenate into a single array {$r_i = \{r_{11}, r_{12}, \dots, r_{1\nsamp}, r_{21}, r_{22}, \dots, r_{2\nsamp} \}$}.  We can write the likelihoods in the presence of a signal, and in the presence of only noise as:
\begin{align}\label{eq:residual-correlation-likelihood}
    p(r|\covgw) = \tr{Det}(2\pi \covgw)^{-1/2} \, \tr{Exp}\lrs{-\frac{1}{2} r_i \, \covgw^{-1}_{ij} \, r_j}, \msp \msp
    p(r|\covnoise) = \tr{Det}(2\pi \covnoise)^{-1/2} \, \tr{Exp}\lrs{-\frac{1}{2} r_i \, \covnoise^{-1}_{ij} \, r_j}.
\end{align}
Here the covariance matrices are, $\covgw_{ij} = \langle (s_i + n_i) \, (s_j + n_j) \rangle$ with GWs, and $\covnoise_{ij} = \langle n_i \, n_j \rangle$ with only noise.  If we again assume that the noise is uncorrelated in time and between pulsars, and further that the GW signal is uncorrelated in time, these matrices will be block diagonal and diagonal respectively,
\begin{align}\label{eq:correlation-covariance-matrices}
    \covgw = \begin{bmatrix*}[c]
        \begin{bmatrix*}[c]
            S_{n_1} + S_s & 0 \\
            0 & \ddots \\
        \end{bmatrix*}
        \begin{bmatrix*}[c]
            S_s & 0 \\
            0 & \ddots \\
        \end{bmatrix*} \\
        \begin{bmatrix*}[c]
            S_s & 0 \\
            0 & \ddots \\
        \end{bmatrix*}
        \begin{bmatrix*}[c]
            S_{n_2} + S_s & 0 \\
            0 & \ddots \\
        \end{bmatrix*} \\
    \end{bmatrix*}, \msp \msp
    \covnoise = \begin{bmatrix*}[c]
        \begin{bmatrix*}[c]
            S_{n_1} & 0 \\
            0 & \ddots \\
        \end{bmatrix*}
        \begin{bmatrix*}[c]
            0 & 0 \\
            0 & \ddots \\
        \end{bmatrix*} \\
        \begin{bmatrix*}[c]
            0 & 0 \\
            0 & \ddots \\
        \end{bmatrix*}
        \begin{bmatrix*}[c]
            S_{n_2} & 0 \\
            0 & \ddots \\
        \end{bmatrix*} \\
    \end{bmatrix*}.
\end{align}
Each of the four `blocks' of these matrices are shaped $\nsamp \times \nsamp$.  The power in the signal and in the noise of each pulsar is: $S_h = \langle s^2 \rangle$, $S_{n_1} = \langle n_1^2 \rangle$, and $S_{n_2} = \langle n_2^2 \rangle$.  The likelihoods can be analytically maximized to find the maximum likelihood estimates of the powers: \mbox{$\hat{S}_h = \hat{S}_{12}$}, \mbox{$\hat{S}_{n_1} = \hat{S}_{11} - \hat{S}_{12}$}, and \mbox{$\hat{S}_{n_2} = \hat{S}_{22} - \hat{S}_{12}$}, where
\begin{align}
    \hat{S}_{12} = \frac{1}{N} \sum_{i=1}^{N} r_{1i} \, r_{2i}, \msp \msp
    \hat{S}_{11} = \frac{1}{N} \sum_{i=1}^{N} r_{1i}^2, \msp \msp
    \hat{S}_{22} = \frac{1}{N} \sum_{i=2}^{N} r_{2i}^2,
\end{align}
We can also analytically determine the maximum-likelihood ratio,
\begin{align}\label{eq:correlation-ml-ratio}
    \Lambda_\tr{ML} \equiv \frac{\tr{Max}\lrs{p(r|\covgw)}}{\tr{Max}\lrs{p(r|\covnoise)}} =
    \lrs[-\nsamp/2]{1 - \frac{\hat{S}_h^2}{\hat{S}_1 \, \hat{S}_2}}.
\end{align}
It is often convenient to define another statistic to be twice the logarithm of the maximum-likelihood ratio which, in the weak signal limit, becomes the SNR of the cross-correlation:
\begin{align}\label{eq:correlation-det-stat}
    \Lambda \equiv 2 \, \ln \lr{\Lambda_\tr{ML}} \approx \frac{N \, \hat{S}_h^2}{\hat{S}_{n_1} \, \hat{S}_{n_2}}.
\end{align}

Typically noise will, however, be correlated in time.  In this case the likelihoods (Eq.~\ref{eq:residual-correlation-likelihood}) cannot be maximized analytically.  However, an optimal statistic can still be calculated numerically and applied directly to sets of TOAs from an arbitrary number of pulsar pairs \citep[see Ch.~7 of][]{Taylor-2021}.  While the noise may be correlated in time, it is typically stationary, such that it is uncorrelated in frequency space.  In that case, a frequency-domain version of Eq.~\ref{eq:cross-correlation} will still hold with, $\langle \rho(f) \rangle = \langle \tilde{r}_1(f) \, \tilde{r}^*_2(f) \rangle \approx \langle \tilde{s}(f) \, \tilde{s}^*(f) \rangle$, where $\tilde{r}^*$ is the complex conjugate of the Fourier transform $\tilde{r}(f) \equiv \mathcal{F}\lr{r(t)}$, given in Eq.~\ref{eq:fourier-transform}.  In this case, the frequency-space covariance matrices look very similar to Eq.~\ref{eq:correlation-covariance-matrices}, and analytic solutions analogous to Eq.~\ref{eq:correlation-ml-ratio} and Eq.~\ref{eq:correlation-det-stat} can again be found \citep[see Sec.~4.3 of][]{Romano+Cornish-2017}.

% {\color{red}JR: I think there are problems with the following paragraph.  I'm not sure how best to rewrite it, given that the OS has three different use-cases in the literature: (i) as a cross-correlation estimator $\hat\rho_{ab}$ for individual pulsar pairs, whose expectation value $\langle\hat\rho_{ab}\rangle=A_{\rm gw}^2 \Gamma_{ab}$; (ii) as a quadratic estimator $\hat A^2_{\rm gw}$ of the squared GW amplitude using all possible pairs of pulsars, and (iii) a SNR detection statistic $\hat\rho \equiv \hat A_{\rm gw}^2/\sigma_0$, where $\sigma^2_0$ is the variance of $\hat A_{\rm gw}^2$ in the absence of a GW signal. (For a detection statistic you always want the variance of your statistic to be calculated in the absence of a signal.) It looks like you are following the presentation from Rosado et al.  If so, then (70) below should be $S_{ijk} =s_i^*(f_k)s_j(f_k)$ without any multiplicative factors or integral.  Also, $S_{h0}(f)$ is usually take to be a {\it normalized} power spectrum, written with an overbar like $\bar{S}_{h0}(f)$ such that $S_{h0}(f)= A_{\rm gw}^2 \bar{S}_{h0}(f)$.  You need to use a normalized power spectrum when constructing a statistic, since you don't know a~priori what the amplitude of the GW signal is; that's usually what your statistic is trying to estimate.  I'd be happy to talk about this some more via zoom if you want.}
We wish to construct an `\term{optimal statistic}' ($\ostat$) defined to maximize the SNR.  Following \citet[][Sec.~III]{Anholm+2009}\footnote{See also \citet{Rosado+2015}, whose notation we employ.}, we define the SNR as \mbox{$\tr{SNR}^2 = \langle \ostat(\tr{GW}) \rangle^2 / \langle \ostat^2(\tr{noise}) \rangle$}, which is the power of the statistic in the presence of GWs, divided by the variance of the statistic in the absence of GWs, such that $\tr{Var}\lr{\ostat(\tr{noise})} = \langle \ostat^2(\tr{noise}) \rangle - \langle \ostat(\tr{noise}) \rangle^2 = \langle \ostat^2(\tr{noise}) \rangle$.  This definition of the SNR is appropriate for weak GW signals, i.e.~when the SNR is small.  The optimal statistic can then be calculated as,
% [Rosado+2015] Eq. A10
\begin{align}\label{eq:optimal_statistic}
    \ostat = \sum_{i=1}^{j} \sum_{j=1}^{N_p} \sum_{k=1}^{\lr{N_f}_{ij}} \frac{\Gamma_{ij} \, \bar{S}_{h0}(f_k)}{P_i(f_k) \, P_j(f_k)} s_{ijk},
\end{align}
where $\Gamma_{ij} = \lr{3/2} \, \hdfunc(\gamma_{ij})$ is a rescaled Hellings-Downs correlation (Eq.~\ref{eq:hd_hd}), for an angle $\gamma_{ij}$ between pulsar $i$ and $j$; and $\bar{S}_{h0}$ is a template spectrum normalized by the (unknown) amplitude of the GWB spectrum, i.e.~$\bar{S}_{h0} = S_{h0} / A^2_\tr{GWB}$.
The summations should be followed from the inside out, where the number of frequency bins for a particular pair of pulsars is $\lr{N_f}_{ij}$, and the number of pulsars is $N_p$.  To simplify the notation, we will replace the triple summation in Eq.~\ref{eq:optimal_statistic} with $\sum_{ijk}$.
% The filtered and noise-weighted cross-correlation between pulsars $i$ and $j$, at a frequency $f_k$ is,
% \begin{align}
%     s_{ijk} = 2 \int_{0}^{+\infty} \delta_T(f_k - f') \, \tilde{s}_i(f_k) \, \tilde{s}_j(f')
%         \frac{S_h(f_k) \, \Gamma_{ij}}{P_i(f_k) \, P_j(f')} \, d f'.
% \end{align}
The cross correlation between pulsars $i$ and $j$, at a frequency $f_k$ is,
% [Rosado+2015] Eq. 21
\begin{align}
    s_{ijk} = \tilde{s}^*_i(f_k) \, \tilde{s}_j(f_k),
\end{align}
where $\tilde{s}^*$ denotes the complex conjugate of the signal in Fourier space.
This optimal statistic is also a `matched filter', comparing the data to a template signal spectrum $S_{h0}$.
% The Fourier transform of the window function, or the approximation to a delta function for a finite time-span $\tobs$, is:
% \begin{align}
%     \delta_T(f) \equiv \frac{\sin\lr{\pi f \tobs}}{\pi f}.
% \end{align}
The mean statistic in the presence of a GW signal is,
\begin{align}
    \langle \ostat(\tr{GW}) \rangle = \sum_{ijk} \frac{\Gamma_{ij}^2 \, S_h(f_k) \, S_{h0}(f_k)}{P_i(f_k) \, P_j(f_k)},
\end{align}
and the variance in the absence of a GW is,
\begin{align}
    \langle \ostat^2(\tr{noise}) \rangle = \sum_{ijk} \frac{\Gamma_{ij}^2 \, S_{h0}^2(f_k)}{P_i(f_k) \, P_j(f_k)}.
\end{align}
For a (correctly) matched filter, i.e.~$S_{h0} = S_h$, the SNR is then $\langle \ostat(\tr{GW}) \rangle^{1/2}$.

% ---- Massive Black Hole Binaries
% ------------------------------------------------------------------------------

\section{Super-Massive Black-Hole Binaries}\label{sec:mbhbs}

\introref{For additional details, we refer the reader to the chapters in this volume on SMBHs (Beckmann \& Smethurst 2025), and SMBH Binaries (Foord 2025).}

As introduced in Sec.~\ref{sec:intro}, \term{super-massive black-hole binaries (SMBHBs)} have long been proposed as sources of the loudest gravitational waves in the Universe.  SMBHs, when accreting sufficient amounts of gas, are visible as `\term{active galactic nuclei (AGN)}', with many quasars (the brightest examples of AGNs) observed from even the very distant and early Universe.  More locally, where the kinematics of gas and/or stars can be resolved in the central $\sim\pc$ of galaxies, dynamical modeling demonstrates the presence of SMBHs in galactic nuclei even when they're inactive.  Observations are consistent with virtually all galaxies hosting an SMBH in their center.  SMBH masses are observed to be tightly correlated with host-galaxy properties \citep{Kormendy+Ho-2013}.  Many empirical relationships have been derived, with the SMBH-mass~vs.~stellar velocity-dispersion relation (\msigma{}) typically regarded as the most precise, but the SMBH-mass~vs.~stellar bulge-mass relation (\mmbulge{}) often being the most convenient.  SMBH masses in the local Universe ($z\approx 0$) are well described by a log-normal distribution with mean and standard deviation (\textit{Ibid.}),
\begin{align}
    \logten{\frac{\mbh}{\msol}} \sim \mathcal{N}\lr{\mu + \alpha_\mu \, \logten{\frac{M_\tr{bulge}}{10^{11} \, \msol}}, \epsilon_\mu} \\
    \mu \approx 8.69 \pm 0.05, \msp \msp \alpha_\mu \approx 1.17 \pm 0.08, \msp \msp \epsilon_\mu \approx 0.28.
\end{align}
An effective heuristic is that SMBH masses are $200$ times less than the galaxy stellar mass\footnote{The fraction of stellar mass in the stellar bulge, the `bulge fraction' $\fbulge$, varies by a factor of a few, but for the high-mass SMBHs of interest to PTAs, $\fbulge \gtrsim 0.7$}, with a standard deviation close to a factor of two in log-space.

Structure formation in the Universe is fundamentally hierarchical in nature, with massive galaxies growing through the merger of many smaller galaxies.  While galaxy mergers are abundantly identified in galaxy surveys, the difficulty of associating a lifetime to those mergers makes the calculation of a galaxy merger-rate non-trivial.  Typical merger rate estimates are either calculated directly from cosmological hydrodynamic simulation, or utilize their merger timescales to normalize observational counts of galaxy pairs.  A comprehensive fit to simulated galaxy merger rates are provided by \citet{Rodriguez-Gomez+2015}.  For parameters near those most important for PTAs, their results can be \textit{very-roughly} approximated as:
\begin{align}
    \frac{d^2 N}{dq \, dz} = 0.92 \, \tr{Gyr}^{-1} \, \scale[0.41]{M}{6\E{11} \, \msol} \, \scale[-1.3]{q}{1/4} \, \scale[2.0]{1+z}{2.0}.
\end{align}
Most massive galaxies then have the opportunity to host multiple pairs of SMBHs over their cosmic histories.

Galaxy mergers bring two SMBHs into a common, post-merger host galaxy, with a separation of $a \sim 10^3~\pc$.  At this time, the mass of material (stars, dark matter, gas) in between the two SMBHs is significantly greater than their own masses, and their direct gravitational influence on each other is negligible.  Based on the typical central densities of massive galaxies, it is only at separations of $a \lesssim 10~\pc$ that the mass of material surrounding each SMBH becomes comparable to the SMBHs themselves.  Once the two SMBHs come closer than this point, their `\term{spheres of influence}', they become gravitationally bound as a true `binary'.  As binaries emit GWs, the system loses energy, and the semi-major axis $a$ decreases.  Recall that it is only at $a \lesssim 1 \, \pc$ before GW-emission alone is sufficient to drive binaries to coalesce: the `coalescence time' $\taulife \equiv \int \lr[-1]{df/dt} df \sim 1 - 10 \, \tr{Gyr}$ (from $\sim$parsec separations; Eqs.~\ref{eq:lifetime_gw_sepa}).  The separations corresponding to PTA-sensitive frequencies are only slightly smaller, $a \lesssim 10^{-2} \, \pc$ (Eq.~\ref{eq:kepler_freq}); and thus, after galaxy merger, binaries need to traverse $\sim$five orders of magnitude in separation to produce detectable GW emission.  Recall also that once in the PTA band, the binary `\term{hardening timescale}' is $\tau \equiv f / \lr{df/dt} \sim 10^5 - 10^6 \, \yr$ (Eqs.~\ref{eq:hard_time_gw_freq}).

These characteristic time scales already tell us a good deal about SMBHB populations.  First, because the time-scale between galaxy mergers is comparable to the time-scale of binary coalescence, a typical massive galaxy can host of order unity dual-/binary- SMBHs at any given time!  Due to the steep lifetime scaling ($\taulife \propto a^4$), there are drastically fewer binaries at smaller separations.  For PTA-band SMBHBs specifically, we expect $\lesssim 10^{-3} \, \fcoal~\tr{galaxy}^{-1}$ hosting binaries in the low-redshift Universe.  Here, $\fcoal = \fcoal(M,q,z,\dots)$ is the coalescing fraction---the fraction of binaries which will coalesce before redshift zero (discussed more below), which is nearly identical to the fraction of binaries reaching PTA-frequencies.  The fraction of AGN containing binaries is more difficult to estimate.  While AGN are known to be triggered by mergers, the difference between typical AGN lifetimes and the delay time between AGN activity and binaries reaching the PTA band is unconstrained\footnote{Fast mergers may preferentially produce binaries in AGN, but it's also possible that binaries reach the PTA band specifically after AGN activity tends to cease.}.  However, if we assume that activation and reaching-the-PTA-band are uncorrelated, and that AGN lifetimes are $10^7 - 10^8~\yr$ \citep[or $\fagn \sim 10^{-2}$, e.g.][and references therein]{Hopkins+2008}, we can estimate that $\lesssim 10^{-5} \, \fcoal~\tr{AGN}^{-1}$ host PTA-band binaries in the low-redshift Universe.

\subsection{SMBH Binary Evolution}\label{sec:binary_evolution}

\introref{For additional details, we refer the reader to \citet{Begelman+1980}, \citet{Yu-2002}, \citet{Milosavljevic+Merrit-2003}, and \citet{Kelley+2017a}.}

A variety of `\term{environmental interactions}' between the binary and components of the host galaxy are required to bring the two SMBHs from galaxy scales to the post-merger galactic nucleus, and then the GW regime.  The process for SMBH binary evolution was first outlined by \citet{Begelman+1980}.  At large separations ($a \sim 10^2 - 10^4 \, \pc$), `dynamical friction' governs the galaxy-galaxy merger itself, and the early inspiral of the two SMBHs in the post-merger host galaxy.  Near the spheres of influence of the SMBHs ($a \sim 10 \, \pc$), where they become gravitationally bound, the dynamical friction formalism breaks down and individual three-body `stellar scattering' events must be considered instead.  In gas-rich mergers, which are more common in lower-mass galaxies, a `\term{circumbinary accretion disk}' can form around the binary at similar scales $a \sim \pc$, and torques from this disk can further influence the binary evolution.  Finally, as shown above, GW emission becomes dominant at the smallest separations ($a \lesssim \pc$) leading to eventual binary coalescence (described in Sec.~\ref{sec:binary_gws}).  The broad picture of this evolution is still believed to hold true, with at least some fraction of binaries able to reach the GW-driven regime, and eventually coalesce.  The details in every phase, however, remain highly uncertain, as are the distribution of typical binary lifetimes and coalescing fractions.

\subsubsection{Dynamical Friction}

In the dynamical friction regime, both SMBHs are initially fully embedded within their host galaxies.  In the early stages, the dominant mass-constituents are the two dark-matter halos, although stars and sometimes gas are non-negligible.  Conceptually, and in simplified calculations, we often consider the secondary SMBH/galaxy inspiraling into the halo/galaxy of the primary.  The standard, idealized model of dynamical friction developed by  \citet{Chandrasekhar-1943} considers a point-mass moving through a uniform, gravitating background that then becomes perturbed.  This model wouldn't seem to hold at all: the two galaxies are comparable in size, and thus fundamentally extended; similarly, there is no uniform background of material; and galaxy mergers are typically highly disruptive.

Oddly then, the simplified formalism of dynamical friction still agrees fairly well with detailed simulations.  Two key considerations are, however, important.  First, accounting for `stripping' of the secondary galaxy: as it inspirals, the furthest-out and least-bound material is gradually removed and deposited into the outer regions of the primary galaxy.  This is induced both by tidal forces and ram-pressure ablation from the head wind that the secondary experiences.  After a few dynamical times, the secondary galaxy will be stripped from the secondary SMBH, leaving only a tightly bound core of stars and gas.  Second, accounting for the multi-component radially-stratified composition of the post-merger galaxy: the density and velocity of the dark matter, stars and gas each change significantly, and independently, with radius which much be modeled.

It is expected that the dynamical friction phase is not always successful in that the secondary SMBH can `\term{stall}' in the outskirts of the primary galaxy, typically at $\sim \kpc$ scales.  This is more common in more-extreme mass-ratio mergers ($q \lesssim 10^{-1}$) and/or less penetrating initial orbits (initial pericenter distances $\gtrsim 1 \, \kpc$), which produce the satellite/dwarf galaxies and tidal streams seen in the Milky Way.  In these cases, the secondary never reaches sufficiently high densities, and/or tidal stripping decreases the secondary mass too quickly.  The resulting `wandering' or `orphan' SMBHs in galaxy outskirts would be difficult to detect or constrain even within the Milky Way as these SMBHs are far less likely to be accreting.  Lensing surveys of the halo, and possibly `tidal disruption events' (TDEs; described below) are likely the best channels.

\subsubsection{Stellar Scattering}\label{sec:binary_evolution_scattering}

Once the energy of the two SMBHs becomes comparable to the local stellar background, typically soon after they form a bound binary, the dynamical friction formalism breaks down\footnote{At this point the `back reaction' on the background can no longer be treated as perturbative.}  Eventually, individual three-body `stellar scattering' events must be considered.  The parameter space of stars that are able to interact with the SMBHB is called the `loss cone' (LC), this includes stars with a range of energies (semi-major axes), and small angular momenta (high eccentricities).  Scattered stars tend to extract energy from the binary, allowing it to continue hardening, while those stars are then ejected from the LC.  For every e-folding of binary SMBH separation, a mass of stars roughly equal to that of the SMBHB must be scattered.  For a $\sim 10^9 \, \msol$ binary, a scattering rate of $\gtrsim 1 - 10 \, \yr^{-1}$ must be sustained for $\sim \tr{Gyr}$ to merge the system.  The population of ejected stars can thus be a substantial fraction of a galaxy's stellar core.  The resulting `core scouring' may be observable, and indeed observed, as the flattened or `cored' stellar density profiles in some massive galaxies.

The efficiency of stellar scattering in merging the SMBHB is determined by two effects: the maximum number of stars in each particular galaxy's LC, and how effectively the LC can be replenished as stars are ejected.  The challenge to refill the LC, and the subsequent stalling of SMBHBs in the stellar scattering regime, is referred to as the `\term{final-parsec problem}' \citep{Milosavljevic+Merritt-2003}.  In idealized models, particularly spherically-symmetric ones, the rate at which LC stars are replenished is very slow: governed by two-body diffusion at the edge of the LC.  In this case, the SMBHB inspiral will often stall.  Over the last decade, a general consensus has emerged that more realistic galaxies have strongly asymmetric, `tri-axial' gravitational potentials (particularly following galaxy mergers) which act to stir the stellar phase-space distribution.  This can rapidly refill the LC to an effectively-full steady-state.  In this case, the stellar scattering rate can remain high, and in many cases the binary will continue to inspiral effectively.

Despite this growing consensus, stellar scattering remains very difficult to model.  Scattering events themselves require resolving $\lesssim pc$ scale sizes, and $\lesssim \yr$ timescales, while the refilling of the loss-cone depends on the galaxy at $\sim \kpc$ sizes and $\gtrsim 10^6 \yr$ timescales.  Even if the LC remains full, the effectiveness of stellar scattering still depends on the central mass, density, and velocity profile of stars which are usually unresolvable in EM observations.  Some fraction, possibly substantial, of binaries could still stall at $\sim \pc$ separations.  If the nanohertz GWB measured by PTAs is confirmed to be produced by SMBHBs, it would be the first direct evidence that the final-parsec problem is indeed solved, at least for a large portion of the most-massive SMBHBs.

\subsubsection{Circumbinary Disks and Torques}

% \needcite{Cross-reference Diego's Circumbinary disk chapter.}
\introref{For additional details, we refer the reader to the chapter in this volume on circumbinary accretion disks (Mu{\~n}oz 2025).}

In accretion disks, internal `viscous' stresses transport angular momentum outwards which allows material to move inwards.  Around compact objects, the viscous stresses are produced by the `magneto-rotational instability' (MRI).  When a second, massive object is added as a binary companion within an accretion disk, three distinct regions form, depicted in Fig.~\ref{fig:cbd-electromagnetics}(a).  At large radii, much larger than the binary separation ($r \gg a$), the `\term{circumbinary disk}' (grey) behaves as if it were around a single-object, now with the combined mass of both components.  At small radii surrounding each component---specifically well-within each's Hill surface ($r \lesssim r_\tr{hill}$)---two `circumsingle disks' behave like perturbed single accretion disks around the primary (red) and secondary (blue).  At intermediate radii ($r_\tr{hill} < r \lesssim 2 a$), orbiting material is no longer stable and a `gap' is cleared.

The overall accretion rate through the circumbinary disk and through both circumsingle disks is relatively unaltered by the presence of the gap.  An over-density tends to form at the inner-edge of the circumbinary disk (Fig.~\ref{fig:cbd-electromagnetics}(a); darker grey), which then overflows and feeds the circumsingle disks through streams.  Even when the binary system begins to inspiral rapidly due to GW emission, more rapidly than viscous torques within the accretion disks can feed material, the accretion continues to be driven by dynamical torques.  Thus, the current consensus is that accretion will continue until nearly the time of coalescence\footnote{At some point, possibly as late as the final few orbits, the circumbinary accretion disk will decouple and no longer be able to feed material as rapidly as the binary inspirals.}, which offers promising opportunities for EM counterparts throughout the inspiral process (Sec.~\ref{sec:ems}).

While the net accretion rate is unchanged by the presence of the binary, the amount of material accreting onto each component can be highly variable.  Specifically, the accretion rate is often periodically modulated at the binary orbital period, and/or the orbital period of the inner-edge of the circumbinary disk (typically $\approx 5-10$ times longer).  Simulations show that accretion tends to favor the lower-mass secondary, at times by large factors.  For some orbital configurations, particularly near-equal-mass binaries, this `preferential accretion' can alternate between the two components on secular timescales of hundreds to thousands of orbits.  The detailed sharing of material seems to be affected by not only the mass-ratio and eccentricity, but also the thermal structure of the disk, all topics of ongoing study.

The biggest surprise in the recent study of circumbinary disks lies in the gravitational interaction between the disk(s) and binary.  It was long believed that circumbinary disks would always extract angular momentum from binaries, hastening their inspiral.  A classic derivation common in planetary dynamics literature is to calculate torques from Lindblad (spiral wave) resonances that result from a first-order expansion of the binary gravitational potential.  Those torques on the binary are negative, extracting angular momentum.  A large number of hydrodynamic simulations, starting with \citet{Miranda+2017}, have now shown that this is not always the case.  The density distribution, particularly in the circumsingle disks, from a full solution to the equations of motion are such that the net torques can be positive: depositing additional angular momentum into the binary, acting against their inspiral \citep{Muñoz+2019}.

The implications of this paradigm shift are still being studied.  As for the partitioning of accretion material, the detailed torque balance appears to depend on the binary parameters and also the thermal structure of the accretion disks.  Both theoretical and observational arguments suggest that in many situations (for example in extreme mass-ratios) the torques should indeed be negative, or otherwise relatively small.  Three-dimensional, non-idealized (e.g.~radiative and/or magnetic) accretion disk simulations are currently only computational feasible for relatively small numbers of orbits.  Similarly, large explorations of parameter space remain intractable.  While the picture is still evolving, there are two reasons to expect that disk `softening' is less important for the highest-mass SMBHBs that PTAs are most sensitive to \citep{Bortolas+2021}.  First, high accretion rates are required for disk densities to become dynamically important, and the highest-mass SMBHs tend to have lower accretion rates.  Second, if such high accretion rates are achieved and maintained, the SMBHBs may grow sufficiently to enter the GW-dominated regime and still coalesce effectively.

\subsubsection{Other Effects}\label{sec:binary_evolution_other}

The most important environmental processes for PTA-detectable SMBHB evolution are dynamical friction and stellar scattering.  Circumbinary disk torques may also play a role, but are likely less important for the highest-mass SMBHBs that PTAs are most sensitive to.  A number of other processes may be non-negligible in the evolution of some SMBHBs.
\begin{itemize}
    \item \textbf{SMBH triple interactions}.  As we've seen, the expected lifetime of SMBHBs is comparable to the time between galaxy mergers.  This means that it may be common for a subsequent galaxy merger to deliver a third SMBH component.  Particularly when the first pair of SMBHs stall, triple interactions could proceed.  Indeed, more careful analysis suggests that up to tens of percent of SMBHBs could be affected by a third SMBH.  In the context of idealized triple interactions, e.g.~as applicable to stellar clusters, the lowest mass component will typically be ejected\footnote{These ejected triples can produce `offset' or `wandering' SMBHs.  See Sec.~\ref{sec:ems_other}.} while hardening a binary of the two heavier components.  It has been proposed that this serves as a mechanism to induce stalled binaries to coalesce and thus establish some minimum level of coalescences, and thus GW signals, regardless of uncertainties in environmental interactions \citep{Ryu+2018, Bonetti+2018}.  These `strong' triple interactions are likely to occur in some fraction of systems, particularly when all three SMBHs have small distances of closest approach.  More common will be `weak' triple interactions at larger separations, where the galaxy potential and dissipative environmental processes are still relevant.  In these cases the dynamics may resemble planetary-like systems.  Kozai-Lidov type interactions could still drive mergers through eccentricity excitation, but in many cases they may be damped by the environment, or simply take longer than a Hubble time.  These systems are particularly challenging to model realistically and require further study.
    \item \textbf{GW recoils.}  GW emission becomes stronger and stronger until binaries coalesce.  During the final orbits, GW emission can become asymmetric, particularly when the SMBHs have large and misaligned spins.  The asymmetric GWs carry momentum, producing a `recoil' in the merged SMBH remnant \citep{Blecha+2016}.  The spin distributions of single SMBHs are extremely uncertain, and those of binaries are completely unconstrained.  For typical assumptions, we expect that most recoils will be $\lesssim 100\tr{s}~\rm{km\ s}^{-1}$, but some fraction can reach $\gtrsim 10^3~\rm{km\ s}^{-1}$, ejecting the remnant from the host galaxy.  Few, if any, massive galaxies seem to be missing their central SMBHs, but no studies have carefully constrained the allowed recoil velocities under these constraints.  It remains unclear whether this implies that SMBHBs stall, that spins are small or aligned, or that new SMBHs can be formed (or brought in) to replace ejected ones.
    \item \textbf{Preferential accretion.}  As discussed above, hydrodynamic simulations show that the smaller secondary component tends to accrete more material than the primary.  In some cases it can be an order of magnitude higher or more.  Additionally, the binary configuration may make it easier to exceed typical Eddington limits as the primary could help to prevent material from being blown off through winds.  Combined with the long timescales that SMBH binaries may spend at $a \lesssim \pc$ scales, this implies that the mass-ratio of the binary, and thus its chirp-mass, could change significantly during inspiral.
\end{itemize}

\subsection{Electromagnetic Counterparts}\label{sec:ems}

\introref{For additional details, we refer the reader to \citet{D'Orazio+Charisi-2023}.}

Gravitational waves intrinsically probe the central, highest density/mass regions.  Electromagnetic (EM) observations, on the other hand, probe the highest temperature gases that, at times, surround those massive objects.  Each \textit{messenger} thus provides different types of information about their sources, and is subject to different selection effects and modeling degeneracies.  Soon after the first GW detections by LIGO was the first multi-messenger event which firmly demonstrated that having both types of signatures revealed tremendously more than than the sum of each type of observation alone.  As millions of SMBHs are already observed as bright EM sources, SMBHBs also present tantalizing opportunities for `\textit{multi-messenger}' astrophysics.

The majority of SMBH growth, particularly at lower masses, is provided by accretion of gas from galactic nuclei.  When the accretion rates are high, gas becomes hotter and more dense, and produces observable EM emission as `\term{active galactic nuclei}' \citep[AGNs;][]{Antonucci-1993, Netzer-2015}.  The emission often spans the EM spectrum, with different types of AGN providing some of the brightest sources of emission in the radio (radio galaxy/quasar, BL Lac objects), optical (quasi-stellar objects, QSOs; quasars; and Seyferts), and X-Ray/$\gamma$-Ray (blazars).  While the detailed structure and dynamics of AGN accretion flows and emission remains surprisingly unclear, they possess a number of key characteristics.  AGN are extremely variable over a wide variety of timescales, in all emission bands.  Variability is apparent at timescales as short as roughly the light-crossing time of the event-horizon $\approx 3.0\E{4}~\sec~\lr{M/3\E{9}~\msol}$, and on timescales as long as observations exist (many decades).  The variability tends to be `red' in character, i.e.~with higher amplitude variations at longer periods, with signs of excess (pseudo-)periodicity.  AGN also produce a wide variety of emission lines at a range of radii in their accretion disks, and also their more extended gas distributions.  As the orbital velocity increases for material nearer the black-hole, the characteristic width of the emission lines becomes broader and broader.  The Narrow Line Region (NLR) is typically at $\sim 100~\pc$; while the broad-line region (BLR) is typically at $\sim 10^{-2}~\pc$ for optical lines, and up-to the inner edge of the accretion disk for X-Ray lines (e.g.~Fe-K-$\alpha$).

\begin{figure}
    \vspace{-10pt}
    \includegraphics[width=\textwidth]{{{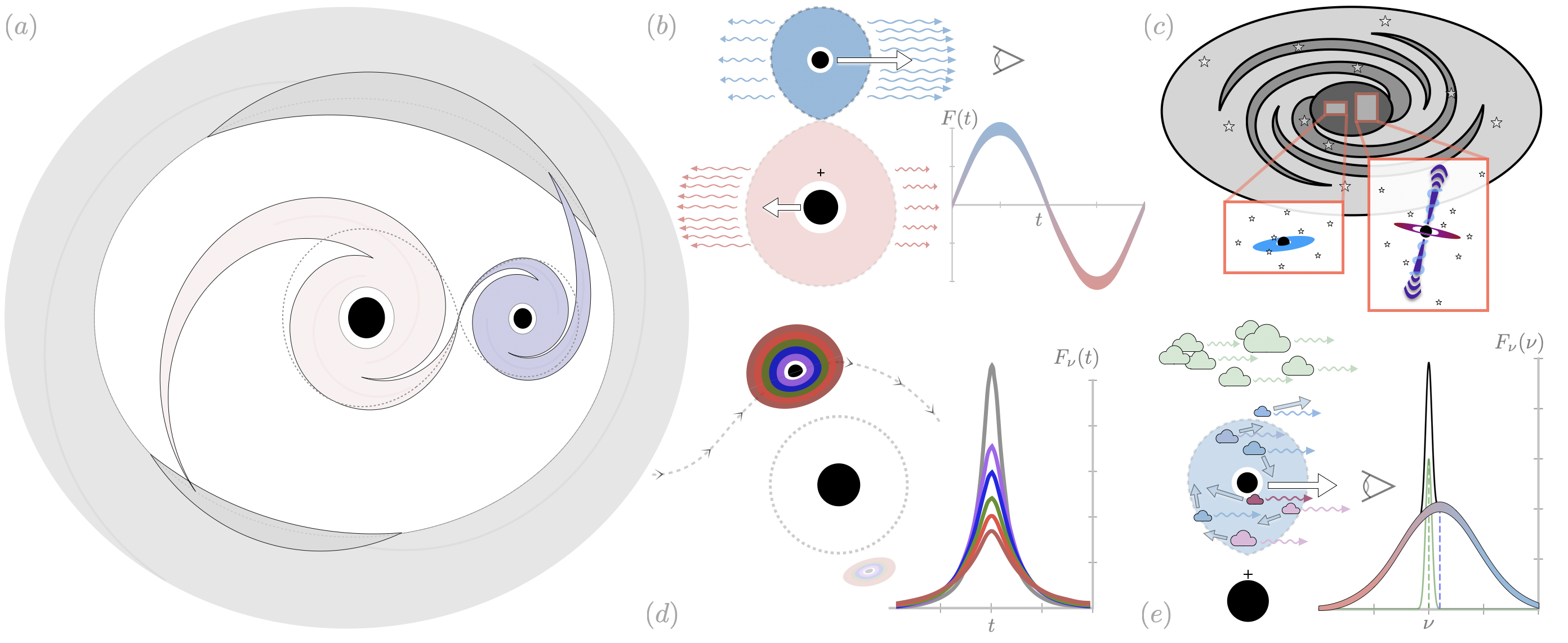}}}
    \caption{\textbf{(a) Accreting binary with circumbinary disk} (CBD; grey), and two circumsingle disks (CSDs; red and blue) truncated at Hill surfaces.  Streams feed CSDs, preferentially the secondary (blue), from over-densities in the CBD.  \textbf{(b) Doppler-boosting} induced periodic photometric-variability of secondary SMBH.  Binaries with orbital planes oriented along the observers line of sight will appear brighter (and bluer) when moving towards the observer, and fainter (and redder) when moving away.  \textbf{(c) Dual AGN} spatially resolved in post-merger galaxy.  Optical and X-ray surveys provide the largest samples, but angular resolution limits most detections to $\sim$kpc separations.  Radio surveys can detected $\sim$pc binaries, but only those in the nearby Universe.  \textbf{(d) Gravitational self-lensing} of secondary with accretion disk, producing lensing flare.  Thermal stratification of the secondary's disk, combined with a lensing radius (grey dotted) comparable in-size to the disk, can produce a chromatic lensing-flare.  \textbf{(e) Kinematic/Spectroscopic offsets} from the broad-lines near the secondary (red-to-blue), relative to narrow-lines surrounding the combined binary at larger distances and lower velocities (green).}
    \label{fig:cbd-electromagnetics}
    \vspace{-10pt}
\end{figure}

A variety of EM signatures have been suggested to signpost the presence of SMBH binaries specifically (discussed below; and shown in Fig.~\ref{fig:cbd-electromagnetics}).  Unfortunately, nearly every EM feature seems to be seen in `normal' individual AGN as well.  A large number of candidate binaries have been identified with these signatures, but significant contamination from false-positives seems necessary.

\subsubsection{Dual AGN}

The only confirmed examples of two SMBHs in the same system are as `dual AGN/SMBHs', where two separate emitting `cores' are spatially distinguishable in a common system, illustrated in Fig.~\ref{fig:gwb}(c).  Most are identified in the X-Ray and optical, and thus restricted to relatively large angular separations, corresponding to physical separations of kpc---long before the two SMBHs become gravitationally-bound as a binary.  These types of observations have clearly demonstrated that SMBHs contained in a merging galaxy are much more likely to be highly accreting than those in isolated galaxies.  This is consistent with theoretical studies and other galaxy merger observations that show large amounts of gas are funneled into galactic nuclei during the galaxy-merger process.  While at first glance this seems promising for the synchronous emission of GWs and bright EM emission, the relative timescales of enhanced accretion and SMBHBs reaching the GW-detectable separations remains unclear.  If the enhanced accretion rates produce bursts of star formation and AGN activity, both of which then result in feedback which then tends to heat up and eject additional gas, it may also be possible that accretion could be suppressed at the time of GW emission.  Post-merger and post-star-burst galaxies are also significantly dustier, which again may make EM observations more difficult.

The closest separation dual-AGN source was identified with radio VLBI observations that resolved a $\sim 10 \, pc$ projected-separation pair of SMBHs \citep{Rodriguez+2006}.  It remains unclear if these are close enough to be gravitationally interacting.  While radio and sub-mm VLBI allows for drastically better angular resolution, only SMBHs in the relatively local Universe are bright enough to observe, significantly limiting the sample of possible host galaxies.  Still, searches for dual, or even binary, SMBHs using Earth-scale interferometers such as the Event Horizon Telescope (EHT) and next-generation very-large array (ngVLA), is an important endeavour.

Ruling out the presence of a non-accreting (or accreting, but spatially unresolvable) SMBH companion to an AGN is surprisingly difficult.  While SMBHBs may produce time-variability signatures (described below), they may also not, and/or those signatures could be on timescales too long to probe.  This, in addition to the difficulty in characterizing selection effects in many dual-AGN surveys, makes the incorporation of dual-AGN observations difficult to incorporate in theoretical studies.

\subsubsection{Periodic Photometric Variability}

Three processes are believed to produce detectable periodic changes in brightness for one or both accreting SMBHs that are specifically in a binary system.
\begin{itemize}
\item \textbf{Binary-induced, hydrodynamic variability.}  Simulations of circumbinary accretion, illustrated in Fig.~\ref{fig:cbd-electromagnetics}(a), show strong periodic modulations of the accretion rate onto one or both components of the binary \citep{Farris+2014}.  Depending on the detailed configuration of the system, the accretion rate can be predominantly modulated on the orbital period: as one of the SMBHs passes near the edge of the circumbinary accretion disk, `grabbing' material; or at the orbital period of the inner-edge of the circumbinary accretion disk itself (a few times the binary orbital period).  If the emission from any of the accretion disks follows the accretion rate changes without too much damping, the overall brightness of the SMBHs can retain some degree of periodic modulation.  Shocks periodically produced within the inner circumsingle or circumbinary accretion disks may also be observably periodic.  More detailed study of the emission produced with variable accretion rates is required to make better predictions for hydrodynamically-induced photometric variability.
\item \textbf{Gravitational self-lensing.}  Binaries oriented nearly edge-on to the observer will have one (or both) components pass in front of the other.  The strong gravitational field of the foreground component can then gravitational lense and significantly brighten the emission from the background component \citep{D'Orazio+DiStefano-2018}, illustrated in Fig.~\ref{fig:cbd-electromagnetics}(d).  Such an event would produce a `lensing flare' with a duration significantly shorter than the binary orbital period and also typically shorter than much of the typical intrinsic-variability of AGN accretion disks, making them easier to identify.  Depending on the detailed configuration of the binary, these lensing events may be chromatic, providing additional means of excluding false positives and additional information about the accretion disk structure.
\item \textbf{Doppler boosting.}  Particularly in unequal-mass binaries, the secondary component can reach mildly-relativistic velocities which can then produce detectable Doppler-induced variations in brightness modulated on the orbital period \citep{D'Orazio+2015}, illustrated in Fig.~\ref{fig:cbd-electromagnetics}(b).  Doppler boosting would again preferentially occur in systems that are nearly edge-on.  Additionally, because both self lensing and Doppler boosting can be directly connected to the binary orbital phase, these two signatures could at times be identified in tandem.
\end{itemize}

\subsubsection{Kinematic/Spectroscopic Offsets}

As described above, the velocity of the SMBHs along the observer's line of sight will produce Doppler boosts.  In addition to increasing and decreasing the brightness of broadband emission, Doppler shifts will change the frequency/wavelength of emitted spectral lines.  Doppler shifted line emission from the circumsingle accretion disks, particularly from the faster-moving secondary, can thus reveal the presence of binary orbital motion.  This is illustrated in Fig.~\ref{fig:cbd-electromagnetics}(e).  The optimal signature would be a broad line that is observed to be time-variable, requiring orbital periods shorter than decades.  Instantaneously-offset broad lines could also signature the presence of a binary, however typical broad lines from single AGN are often asymmetric, offset, and time-variable.  As for periodic photometric variability this likely explains most of the existing spectroscopically-identified binary candidates.  Again, how to filter out false positives remains unclear and an important direction of ongoing research.

Dynamical considerations determine whether a given orbital configuration can plausibly be detected by spectroscopic offsets \citep{Kelley+2021}.  First, the broad-line emitting region must be located at radii that are within the Hill sphere of one or the other binary component.  Second, the projected velocity-offset must be larger than the accuracy by which the emission line `center' can be identified.  Instrumentally, this typically requires $v_\textrm{orb} \gtrsim 10^3 \textrm{ km s}^{-1}$.  These two criteria already significantly restrict the viable parameter space of detectable binaries, such that only a small fraction of all binaries would be detectable.  Much more challenging, however, is the intrinsic variability of normal AGN broad-lines, both in time and across populations. This implies that a selection criterion closer to $v_\textrm{orb} \gtrsim 10^4 \textrm{ km s}^{-1}$, and/or requiring time-changing offsets consistent with orbital motion, should be required.  Either more stringent criteria makes detections fairly unlikely.  Shifted X-Ray broad lines offer a much more promising portion of parameter space, however X-Ray spectroscopic instruments are much more limited in their resolution and sensitivity.

\subsubsection{Other possible signatures}\label{sec:ems_other}

A number of other dual/binary signatures have been proposed in the literature which are worth mentioning.
\begin{itemize}
    \item \textbf{Chromatic Deficits}.  The presence of a binary within an accretion disk clears out a `gap' of material, see Fig.~\ref{fig:cbd-electromagnetics}(a).  Due to the temperature stratification of accretion disks, this suggests a corresponding deficit of emission at temperatures corresponding to the binary separation.  Such signatures could be masked by: differing thermal structures in the circumbinary vs.~circumsingle disks, emission from outside of the plane of the disk, and the presence of general non-thermal emission.
    \item \textbf{Enhanced and double TDEs.}  Tidal-disruption events (TDEs) occur when nuclear stars pass within the tidal disruption radius of a black hole.  Similar to the stellar-scattering hardening mechanism of binary orbits, the population of stars able to make close-encounter orbits can be depleted over time.  However, the presence of a binary, particularly at $\sim \pc$ scales, may replenish or otherwise enhance the rate of encounters.  This has been proposed as a method of explaining the excess of TDE host galaxies showing signs of recent bursts of star-formation and/or galaxy mergers (so-called `E+A' galaxies).  TDE hosts may then be higher-likelihood sources for additional followup in search of additional binary signatures.  In some rare cases, stars in TDEs (or their debris streams) may actually interact-with, or feed, both SMBH components of a binary.  If this produces observable signatures, it could provide a direct indication of a binary.  Post-merger SMBH remnants that experience GW recoils could also experience enhanced TDE rates, offering a probe of post-merger systems.
    \item \textbf{Varsitometry.} The combination of a variable-brightness point-source (i.e.~AGN), and a steady circularly-symmetric source (i.e.~galaxy), leads to a time varying astrometric centroid of the combined light, even if the two sources are well within the point-spread function of the observatory \citep{Shen+2019}.  Differentiating the two components chromatically and/or spectroscopically, along with carefully modeling the variability, allows for the measurement of an offset between the two sources that (to first order) is insensitive to the size of the point-spread function.  For typical sensitivities and configurations this could allow dual AGN to be detected at separations as low as $10s \, \pc$ in the optical, perhaps marginally entering the `binary' regime.  Recently-merged galaxies are often highly asymmetric, making this approach more challenging.
    \item \textbf{Offset/Ejected AGN.}  Just as dual AGN can be spatially resolved at $\sim \kpc$ separations, an individual accreting SMBH may be observably offset from the nucleus/centroid of the overall galaxy.  This could happen during the early inspiral when only one component is active (again, recently-merged galaxies pose a challenge).  As discussed above (Sec.~\ref{sec:binary_evolution_other}), three-body interactions may occur in a noticeable fraction of SMBHBs, in which case one of the components (typically the least massive) can be dynamically ejected from the system.  Additionally, following a binary merger, the combined remnant SMBH can receive a recoil `kick' due to the asymmetric emission of GWs.  In many cases this could produce an offset AGN that wanders the galaxy until it sinks back to the nucleus due to dynamical processes.  In some cases that kick velocity can exceed the escape velocity of the galaxy, leading to SMBHs that are entirely ejected from their host galaxies.  Offset and especially ejected SMBHs will fairly quickly deplete their reservoirs of gas, and are less likely to continue receiving large inflows, so they will preferentially be dark after some period of time.  Even within the Milky Way, identifying an inactive, wandering SMBH would be challenging.
    % Lensing surveys, however ones not directed towards the galactic centers, may be able to detect such systems.
\end{itemize}

% ---- Cosmological Sources
% ------------------------------------------------------------------------------

\section{Non-Binary GW Sources}\label{sec:cosmo}

The standard model of particle physics has proven incredibly successful, but it does not explain a number of fundamental problems, for example: the baryon asymmetry, and the nature of dark matter and dark energy.  For these reasons, it is necessary for there to be additional physics `Beyond the Standard Model' (BSM).  Similarly, a quantum theory of gravity is broadly expected, but has yet to be expounded.  Many BSM models have been proposed, focusing on the behavior of the Universe at very early times and at very large scales.  Many of these models, such as cosmic inflation, generically predict the generation of GWs in the early Universe \citep{Maggiore-2000}.  BSM scalar fields could exhibit phase transitions in the early Universe\footnote{Note that quantum chromo-dynamics from the standard model, also predicts an early Universe phase transition that could produce GWs \citep{Witten-1984}.} which directly, or via topological defects (such as cosmic strings and domain walls), would produce GWs \citep{Kibble-1976}.  The imprints from GWs at ultra-low frequencies (with wavelengths comparable to the observable Universe) in the cosmic microwave background are actively being searched for, typically in the form of `B-mode' polarization signals \citep{Hu+White-1997}.

The recent evidence for a stochastic GWB has also been analyzed in terms of BSM models, and the predictions from a number of models are found to be consistent with current observations \citep{Afzal+2023}.  It is difficult to assess the \textit{a priori} plausibility of these models as their parameter spaces are often entirely unconstrained, and most models haven't been entirely reconciled with typical cosmological/astronomical observations and/or the Standard Model.  At the same time, no confirmed SMBHBs have been identified either in EM surveys or GW observations.  The identification of individual SMBHBs, particularly through GW emission, would strongly suggest they produce the bulk of the GWB.  It is of course possible that both SMBHBs and cosmological sources could both contribute, even if not at an equal level.  A general feature of BSM GW predictions is that the effective number of GW sources is far larger than the expected number of GW emitting SMBH binaries, and their emission is in the very early Universe.  Both factors suggest that a GWB from cosmological/BSM sources would be far more isotropic on the sky than SMBHBs.  This motivates the search for anisotropy as a `smoking gun' for a binary origin to the GWB, whereas stringent limits on the maximum allowable anisotropy could strongly indicate a BSM origin.

\section*{Conclusions \& Outlook}

Pulsar timing arrays (PTAs) have found compelling evidence for a low-frequency (nanohertz), stochastic gravitational-wave background (GWB).  Such a signal has long been proposed from the ensemble of GWs produced by many super-massive black-hole binaries (SMBHBs) distributed throughout the Universe.  While the idealized prediction is a power-law GWB characteristic-strain spectrum with an index of $-2/3$, realistic GWB spectra are likely to deviate significantly due to the discreteness of emitters and non-GW evolution due to environmental interactions.  In fact, these deviations from the power-law behavior encode a wealth of information about SMBHB populations and their evolution.

Current data is unable to confirm whether the measured signal is produced by SMBHBs.  It is also possible that non-standard-model physics in the early Universe could be producing the measured GW signal.  Additional data is needed to definitively establish the origin of the GWB.  Specifically, the identification of individual, loud binaries producing continuous wave (CW) emission, or similarly significant variations in loudness of the GWB either between different frequency bins or different parts of the sky, would definitively establish SMBHBs as the origin of the signal.  In this case, a wealth of possible electromagnetic (EM) counterpart signals could allow for multi-messenger astrophysics.  Identifying EM signals that uniquely identify SMBHBs, as opposed to peculiar normal/individual active galactic nuclei (AGN), remains very challenging.

Over the next few years, PTAs are poised to either (i) answer long-standing questions about the formation and evolution of the Universe's most behemoth SMBH binaries, and/or (ii) measure/constrain beyond-the-standard-model physics in the very-early Universe.  The hunt is on to increase PTA sensitivity, better characterize the GWB spectrum, and measure the contributions to the GWB from different parts of the sky---determining the true source of these GWs and possibly identifying individually-loud SMBHBs, their host galaxies, and their EM counterparts.

% ---- Other Low-Frequency Detections
% ------------------------------------------------------------------------------

% ==============================================================================
% ====    End Matter   ====
% ==============================================================================

\begin{ack}[Acknowledgments]
%! --- NEED A BLANK LINE HERE (without comment)

Special thanks to Joseph Romano and Jeff Andrews for invaluable comments on drafts of this article, and to Bruce Allen for elucidating conversations and very patient explanations.
\end{ack}

\bibliographystyle{aasjournal}
\bibliography{refs}

\end{document}